\documentclass[a4paper,11pt]{article}
\usepackage{jheppub}
\usepackage{amssymb}
\usepackage{amsmath}
\usepackage{mathtools}
\usepackage{amsfonts}
\usepackage{dsfont}
\usepackage{bm}
\usepackage{braket}
\usepackage{bm}
\usepackage{lmodern}

\usepackage{mathrsfs}
\usepackage{tikz}
\usetikzlibrary{shapes}
\usetikzlibrary{calc}
\usetikzlibrary{arrows.meta}
\usetikzlibrary{positioning}
\usetikzlibrary{decorations.markings}
\usetikzlibrary{decorations.pathmorphing}
\usetikzlibrary{fadings}

\usepackage[most]{tcolorbox}
\newtcolorbox{verticallines}[1][]{blanker, breakable, left=3mm, right=3mm, top=1mm, bottom=1mm, borderline vertical={1.5pt}{0pt}{#1}, parbox=false}

\usepackage{comment}

\numberwithin{equation}{section}

\DeclareMathAlphabet{\mathpzc}{OT1}{pzc}{m}{it}

\newcommand{\mrm}[1]{\mathrm{#1}}
\newcommand{\mcl}[1]{\mathcal{#1}}
\newcommand{\mbb}[1]{\mathbb{#1}}
\newcommand{\mbf}[1]{\mathbf{#1}}
\newcommand{\mfr}[1]{\mathfrak{#1}}

\newcommand{\cint}[1]{\underset{#1}{\oint}}
\newcommand{\llangle}{\left\langle}
\newcommand{\rrangle}{\right\rangle}
\newcommand{\br}[1]{\left({#1}\right)}
\newcommand{\bbr}[1]{\big({#1}\big)}
\newcommand{\corr}[1]{\left\langle{#1}\right\rangle}

\title{Super Covering Maps}
\author[a]{Beat Nairz}
\affiliation[a]{Institut f\"ur Theoretische Physik,
ETH Z\"urich,\\
Wolfgang-Pauli-Strasse 27,
8093 Z\"urich, Switzerland}
\emailAdd{nairzb@ethz.ch}

\allowdisplaybreaks[2]

\abstract{We define analytic maps between super Riemann surfaces which extend the notion of branched covering maps to a supersymmetric setting. We show that these super covering maps appear naturally both in symmetric product orbifolds of superconformal field theories, as well as in the hybrid formalism for tensionless string theory on $\text{AdS}_3\times S^3\times\mbb{T}^4$. In the former, they can be used to calculate correlators in a manifestly supersymmetric way, while in the latter they solve Ward identities of worldsheet correlators.}

\begin{document}

\maketitle

\newpage

\section{Introduction}

The duality of string theory on $\text{AdS}_3\times S^3\times \mbb{T}^4$ with pure NS-NS flux and the D1-D5 CFT is among the best studied examples of the AdS/CFT correspondence.
Strings in this background can be described by a WZW model \cite{Maldacena:2000hw} or in the hybrid formalism \cite{Berkovits:1999im}, giving strong control over the worldsheet theory. This has allowed for extensive checks and even proofs that string theory in these backgrounds is dual to deformations of a symmetric product orbifold theory \cite{Eberhardt:2018ouy,Eberhardt:2019ywk,Dei:2020zui,Eberhardt:2021vsx}.

Particularly in the tensionless limit of $k=1$ units of NS-NS flux, dual to the symmetric product orbifold $\text{Sym}^N(\mbb{T}^4)$, a beautiful geometric picture emerges, in which string worldsheets are identified with the covering surfaces appearing in orbifold calculations.\footnote{A similar identification can also be carried out for long strings at $k>1$, see for example \cite{Dei:2022pkr,Knighton:2024qxd,Knighton:2024pqh}. Furthermore, similar geometric identifications appear in a twistor formulation of $\text{AdS}_5$ \cite{Bhat:2021dez}.} This microscopic understanding of the mechanism underlying the holographic duality for tensionless strings demonstrates the importance of the $\text{AdS}_3/\text{CFT}_2$ correspondence as a prototype of holography.

Both sides of the duality are supersymmetric, but this fact is not directly relevant for the Ward identity analysis showing the localisation to covering maps. Indeed, it was recently shown that one can formulate a purely bosonic WZW model that consistently localises \cite{Eberhardt:2025sbi}. Still, calculations in the supersymmetric theory are of interest to flesh out the understanding of the correspondence, see \cite{Iguri:2022pbp,McStay:2023thk,Dei:2023ivl,Sriprachyakul:2024gyl,Naderi:2024wqx,Yu:2024kxr,Yu:2025qnw} for some recent progress. It is thus interesting to better understand the role of the fermions in the correspondence and see if a similar geometric picture of the mechanism can serve to organise the supersymmetric calculations.

In this paper, we describe a partial geometric interpretation of the duality in the manifestly supersymmetric case. We define \textit{super covering maps}, which are analytic maps between super Riemann surfaces. As in the bosonic case \cite{Lunin:2000yv,Lunin:2001pw}, we show that these maps can naturally be used to calculate correlators in symmetric product orbifolds of superconformal field theories. We then also study the Ward identities of string theory correlators. We focus on the case with worldsheet supersymmetry in the RNS formalism and the case with spacetime supersymmetry in the hybrid formalism of \cite{Dei:2020zui}. We show that super covering maps provide natural solutions to these constraints. This extends the geometric understanding of the duality gained in \cite{Eberhardt:2019ywk,Dei:2020zui} to a supersymmetric setting.

\subsection{Covering maps in the \texorpdfstring{$\text{AdS}_3/\text{CFT}_2$}{AdS3/CFT2} correspondence}\label{ssec:cov_maps_in_adscft}

A breakthrough in understanding the $\text{AdS}_3/\text{CFT}_2$ correspondence for the tensionless string was an insight about the role of (branched) covering maps \cite{Pakman:2009zz,Eberhardt:2019ywk,Dei:2020zui}. To serve as a guiding motivation behind the definitions and examples shown in the following sections, we briefly review these objects and the intuition behind their appearance in the correspondence.

Let us first review the CFT side of the story, which is the symmetric product orbifold $\text{Sym}^N\bbr{\mbb{T}^4}$, see for example \cite{Lunin:2000yv,Lunin:2001pw,Pakman:2009zz,Roumpedakis:2018tdb}. For a seed CFT $X$, its symmetric product orbifold theory $\text{Sym}^N\bbr{X} = X^N/S_N$ is obtained by taking $N$ copies of the seed CFT and identifying them using elements of the permutation group $S_N$. Explicitly, let $v$ be a field in the seed $X$ and denote by $v^{(k)}$ the field in the $k$-th copy. The action of the $(1\dots w)$ permutation is implemented by a twist operator $\sigma_w$.\footnote{Due to the identification of the copies, one needs to sum over conjugacy classes of permutations to obtain a gauge invariant operator \cite{Lunin:2000yv}. There is thus a unique gauge-invariant operator $\sigma_w$ corresponding to the conjugacy class of $w$-cycles.} The fields $v^{(k)}$, $k=1,\dots,w$, are permuted when they move around the twist operator,
\begin{equation}
v^{(k)}\bbr{\gamma(1)}\,\sigma_w(0) = v^{(k+1)}\bbr{\gamma(0)}\,\sigma_w(0)\ ,
\end{equation}
where $\gamma(t)= \varepsilon\,e^{2\pi i t}$ is a small circle around $0$. The fields thus become multi-valued in the presence of $\sigma_w$.

Correlators of twist fields can be calculated using the covering space method of \cite{Lunin:2000yv}. On the covering surface $\Sigma$, the twisted boundary conditions are geometrically encoded. Concretely, there is a holomorphic map $\Gamma:\,\Sigma \longrightarrow \mbb{CP}^1$ determined by the field configuration. If $\sigma_{w}$ is inserted at $x_1$, there is one pre-image $z_1$ at which $\Gamma$ has ramification index $w$, i.e.~such that near $z_1$
\begin{equation}\label{eq:usual_covering}
\Gamma(z) = x_1+a\,(z-z_1)^{w}+\mcl{O}\big((z-z_1)^{w+1}\big)\ .
\end{equation}
Everywhere else, $\Gamma$ must be unramified, meaning $\partial \Gamma\neq 0$. This implies that $\Gamma$ is locally a biholomorphism except at the ramification points.

\begin{center}
\begin{minipage}[t]{.4\textwidth}
\vspace{0cm}
\includegraphics[width=0.95\textwidth]{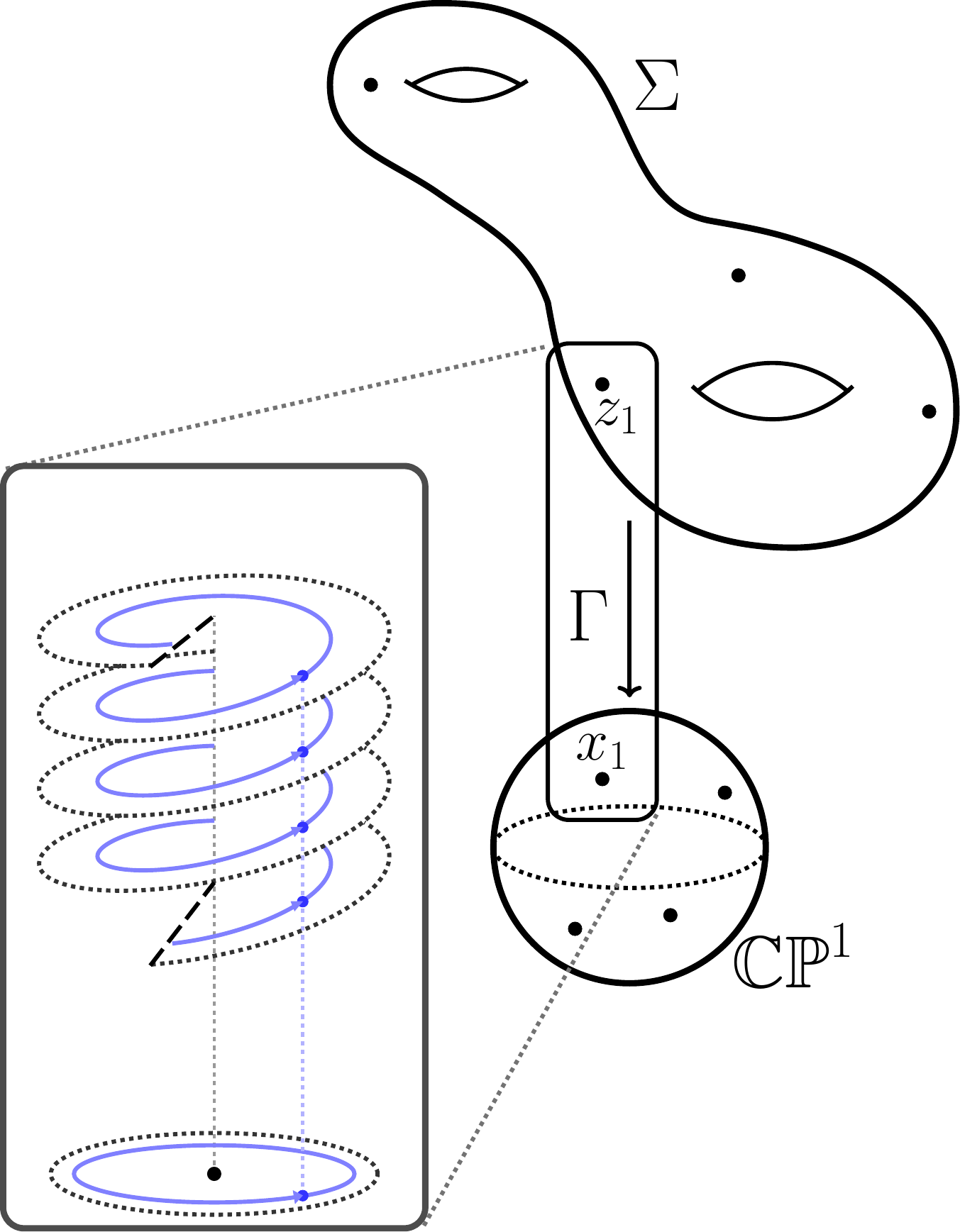}\hfill
\end{minipage}
\begin{minipage}[t]{.59\textwidth}
\vspace{0cm}
The structure of $\Gamma$ is shown schematically on the left. On the covering surface $\Sigma$, there are isolated ramification points $z_i$, which are the ramified pre-images of the branch points $x_i$ on the base space $\mbb{CP}^1$. Zooming in on a ramification point $z_1$, one can see how $\Gamma$ implements the twist. Over $x_1$, the surface $\Sigma$ looks like a $w$ storey car park; moving around $x_1$ corresponds to moving up only one storey but not to a closed contour. Thinking of the label of the copy $k$ as the storey, one can see how the fields $v^{(k)}$ mixed by the twist operator are lifted to a single field $\mcl{V}$ on the covering surface. This is the basic idea behind the application of Hurwitz theory to the symmetric orbifold, which shows how the $S_N$ structure of twist operators is in one-to-one correspondence with covering maps \cite{Cavalieri_Miles_2016}.
\end{minipage}
\end{center}

Let us now briefly mention how the covering surface appears in string theory. There is a nice semi-classical picture which holds for the tensionless string \cite{Eberhardt:2019ywk}. Strings in $\text{AdS}_3$ can become very large --- in fact, it turns out that in the tensionless case they are always as large as they can be, and stretch all the way to the $\mbb{CP}^1$ conformal boundary of $\text{AdS}_3$. The string is thus described by a map from the worldsheet $\Sigma$ to the conformal boundary $\mbb{CP}^1$, and its left-moving degrees of freedom therefore define a covering map $\Gamma$. This covering map can then be directly identified with the one in the dual CFT, and combining it with the right-movers one reproduces the CFT correlator. Technically, the localisation of the string to points in moduli space where the covering map exists is shown using Ward identities \cite{Eberhardt:2019ywk, Dei:2020zui} or in the path-integral formalism \cite{Hikida:2020kil,Dei:2023ivl,Knighton:2024qxd}.

\subsection{Overview}

Does a similar geometric identification hold in the supersymmetric setting? To answer this question, we need to formulate the theories in a manifestly supersymmetric way. We review the relevant notions for two-dimensional CFTs, such as super Riemann surfaces and superconformal structures, in Section \ref{sec:SRS}. Super covering maps will then be functions of an even variable $z$, as well as an odd variable $\Theta$. We define these maps, study some of their properties, and show examples in Section \ref{sec:super_covering_maps}. We can see that these are the correct objects both for the symmetric orbifold, as well as for string theory. We study the CFT side in Section \ref{sec:orbifold}, where we look at the symmetric product orbifold of a superconformal field theory, defined on superspace. The super covering map directly computes correlators in this theory in a manifestly supersymmetric way. We discuss the super Weyl anomaly and the lifting of fractional modes. In Section \ref{sec:string_theory}, we investigate the role of super covering maps in string theory. We first discuss a general picture, as well as several technical complications when combining worldsheet and spacetime supersymmetry. We then look at the string dual in the RNS and hybrid formalism. We define the vertex operators dual to the super vertex operators in the CFT. We analyse the Ward identities and show that they are solved by super covering maps in limits where we keep only either worldsheet or spacetime supersymmetry manifest. For the RNS formalism in Section~\ref{ssec:RNS}, we have manifest worldsheet supersymmetry. In the hybrid formalism analysis presented in Section~\ref{ssec:hybrid}, we instead have manifest spacetime supersymmetry.

\section{Super Riemann surfaces}\label{sec:SRS}

We begin by reviewing the notion of a super Riemann surface (SRS) and a superconformal field theory (SCFT) defined thereon. A SRS is a supermanifold with an additional structure linking the even and odd variables \cite{Friedan:1986rx, DHoker:1988pdl, Witten:2012ga, Witten:2019ylx,Varadarajan:2004yz}. On such a surface, there is an analogue of the usual complex analysis, allowing for differentiation, integration, and Taylor expansion. The superconformal structure is the natural geometric foundation for a SCFT, where a field and its superpartner are combined into a single superfield.

\subsection{Definition}\label{ssec:SRS_defn}

The natural geometric objects in the study of $\mcl{N}=1$ SUSY are super Riemann surfaces (SRS). These are complex supermanifolds \cite{Varadarajan:2004yz} of complex dimension $1|1$, such that their underlying ordinary manifold  $\Sigma^\text{red}$ is a Riemann surface. A SRS is endowed with an additional superconformal structure \cite{Witten:2012ga}. This structure can be abstractly specified as follows. For a SRS $\Sigma$, a \textit{superconformal structure} is a $0|1$-dimensional, maximally non-integrable subbundle $\mcl{D}$ of the tangent bundle $T\Sigma$.\footnote{This is similar to the definition of a complex structure for normal manifolds. There, the tangent bundle of a two-dimensional real manifold is split into a holomorphic and an antiholomorphic subbundle \cite{moroianu_2007}.} Here, ``maximally non-integrable'' means that $[\mcl{D},\mcl{D}]$\ \footnote{The bracket $[\cdot,\cdot]$ denotes the graded Lie bracket on the tangent bundle.} is everywhere linearly independent of $\mcl{D}$.

Concretely, this means that around each point there exist local coordinates $(x,\theta)$, such that $\mcl{D}$ is spanned by the \textit{superderivative}
\begin{equation}
D =\frac{\partial}{\partial \theta} + \theta\,\frac{\partial}{\partial x}\ ,
\end{equation}
which satisfies $D^2 = \partial/\partial_x$ \cite{Witten:2012ga}. Coordinates with this property are called \textit{superconformal coordinates}, and a SRS is defined by an atlas of superconformal coordinate systems. Different coordinates $(\widetilde{x},\widetilde{\theta})$ are superconformal if and only if the transition functions satisfy the following important property
\begin{equation}\label{eq:superconformality_condition}
\widetilde{\theta}\,D\widetilde{\theta} = D\widetilde{x}\ ,
\end{equation}
where $D$ is the superderivative of the $(x,\theta)$ coordinate system.
The most general form of a superconformal transformation is
\begin{equation}
\tilde{x} = f + \theta\,\rho\,\sqrt{\partial f}\ ,\qquad \tilde{\theta} = \rho + \theta\,\sqrt{\partial f + \rho\,\partial \rho}\ ,
\end{equation}
where $f(x)$ is an even, and $\rho(x)$ an odd function.

A SRS can be constructed by gluing together patches of $\mbb{C}^{1|1}$ with superconformal transition functions of the form shown above, as discussed in \cite{Witten:2012ga}. If the odd function $\rho$ in these transitions is zero, $\theta$ transforms as $dz^{\frac{1}{2}}$ under reparametrisations of $z$, i.e.~as an odd section of a square root of the cotangent bundle. A SRS thus includes the choice of a spin structure on the underlying Riemann surface. A SRS for which the transition functions can be chosen to have no odd $\rho$ is called ``split''. In the generic case, the space in which the odd moduli lie also depends on a choice of spin structure \cite[Section 2]{Witten:2012ga}.

\subsubsection{The super sphere}

There is a unique SRS $\mbb{CP}^{1|1}$ whose reduced manifold is the sphere $\mbb{CP}^1$ \cite{Witten:2012ga}. This SRS is specified by two superconformal coordinate patches $U_i=\mbb{C}^{1|1}$ with coordinates $(x_i,\theta_i)$, glued
together with the transition function
\begin{equation}
\big(x_2,\theta_2\big) = \br{-\frac{1}{x_1}, \frac{\theta_1}{x_1}}\ .
\end{equation}
The change of coordinates is superconformal, as
\begin{equation}
D_1 x_2 = \frac{\theta_1}{x_1^2} = \theta_2\,D_1\theta_2\ .
\end{equation}
The underlying ordinary transition function between the even coordinates $x_i$ is just that of $\mbb{CP}^1$.

The automorphism group of $\mbb{CP}^{1|1}$ are the super M\"obius transformations $\text{OSp}(1|2)$ \cite{DHoker:1988pdl}. The general form of such a transformation is
\begin{align}
\widetilde{x}&=\frac{ax+b+\alpha\theta}{cx+d+\beta\theta}\ ,\\
\widetilde{\theta}&= \frac{\gamma x+\delta + A\theta}{cx+d+\beta\theta}\ .\nonumber
\end{align}
Latin letters denote even variables, while greek lettters denote odd ones. The superconformality condition $D\tilde{x} = \tilde{\theta}\,D\tilde{\theta}$ fixes
\begin{equation}
ad-bc = 1+\alpha\,\beta\ ,\quad
\gamma = a\,\beta - c\,\alpha\ ,\quad
\delta = b\,\beta - d\,\alpha\ ,\quad
A = 1-\alpha\,\beta\ .
\end{equation}
We see that there are three even parameters and two odd ones, allowing us to fix three even and two odd coordinates. The two odd generators of the superalgebra $\mfr{osp}(1|2)$ are supertranslations
\begin{equation}\label{eq:super_translation}
\widetilde{x} = x - \lambda\theta\ ,\quad \widetilde{\theta} = \theta + \lambda\ ,
\end{equation}
as well as superscalings
\begin{equation}\label{eq:super_scaling}
\widetilde{x} = x - \kappa\theta x\ ,\quad \widetilde{\theta} = \theta + \kappa x\ .
\end{equation}
For example, if we have three punctures $\mbf{x}_i=\bbr{x_i,\theta_i}$, we can always set them to
\begin{equation}\label{eq:reference_points}
\mbf{x}_1 = \bbr{\infty,0}\ ,\quad \mbf{x}_2 = \bbr{1,0}\ ,\quad \mbf{x}_3 = \bbr{0,\theta_3}\ ,
\end{equation}
which we will use often throughout the paper. Explicitly, this can be done by first applying a supertranslation and superscaling
\begin{equation}
\tilde{x} = x+\theta\,\bbr{\lambda + \kappa\,x}\ ,\tilde{\theta} = \theta+ \bbr{\lambda + \kappa\,x} \ ,
\end{equation}
with
\begin{equation}
\kappa = -\frac{\theta_2-\theta_1}{x_2-x_1} \ ,\qquad \lambda = -\theta_1 - \kappa\,x_1\ ,
\end{equation}
to set the odd components of the first two points to zero. Then, one can apply a usual M\"obius transformation to set the even components to $\infty,1,0$, respectively.

This is the reason for the moduli space of genus zero SRS with $n$ NS punctures\footnote{See \cite{Witten:2012ga} for a discussion on the difference of so-called NS and R punctures. We will comment on R punctures in Section \ref{ssec:R_punctures} and give more details in Appendix \ref{app:R_cov_map}.} having dimension $n-3|n-2$ \cite{Witten:2012ga}.

\subsection{Integration and Taylor expansion}
There is a natural analogue of complex analysis on SRSs \cite{Friedan:1986rx}. In particular, superanalytic functions can be Taylor expanded and integrated over contours. First, we introduce some useful notation to shorten expressions. We denote coordinates $(x,\theta)$ by boldface letters $\mbf{x}$. For two punctures $\mbf{x}_i$, the superdifferences $\mbf{x}_{12}$ and $\theta_{12}$ are defined by
\begin{align}
\mbf{x}_{12}&=x_1-x_2-\theta_1\theta_2\ ,\\
\theta_{12}&=\theta_1-\theta_2\ .\nonumber
\end{align}
These superdifferences are invariant under the supertranslations in eq.~(\ref{eq:super_translation}).

The superanalytic functions on SRSs we consider are $f(\mbf{x}) = f_0(x)+\theta\,f_1(x)$ with $f_0$ and $f_1$ both analytic and with opposite statistics.
The component-wise expansion of $f$ can be repackaged elegantly. Concretely, the Taylor expansion of $f(\mbf{x}_1)$ ``around'' $\mbf{x}_0$ is given by \cite{Friedan:1986rx}
\begin{equation}\label{eq:super_Taylor}
f(\mbf{x}_1) = \sum_{n=0}^{\infty} \frac{1}{n!}\,\mbf{x}_{10}^n\cdot\partial^n\,\big(1+\theta_{10}\,D\big)\big|_{\mbf{x}=\mbf{x}_0}f(\mbf{x})=f(\mbf{x}_0)+\theta_{10}\,Df(\mbf{x}_0)+\mbf{x}_{10}\,\partial f(\mbf{x}_0)+\dots\ ,
\end{equation}
where $\partial = \frac{\partial}{\partial x}$.

Furthermore, we can define contour integration $\oint d\mbf{x}$ on SRS by a combination of Berezin integration over the odd variable and ordinary contour integration. Specifically, if we denote by $\mbf{C}(\mbf{x}_0)$ a supercontour, i.e.~a submanifold of real dimension $1|1$ such that its reduced manifold $C(x_0)$ is a closed curve around $x_0$, then
\begin{equation}\label{eq:super_contour_integration}
\underset{\mbf{C}(\mbf{x}_0)}{\oint}\!\!d\mbf{x}\,\,f(\mbf{x}) := \underset{C(x_0)}{\oint}\!\!dx\,\,\int \!\!d\theta\,\,f(\mbf{x}) = \underset{C(x_0)}{\oint}\!\!dx\,\,f_1(x)\ ,
\end{equation}
where we carried out the Berezin integral in the last step. Under a superconformal change of coordinates, the superdifferential $d\mbf{x}$ transforms as
\begin{equation}
d\widetilde{\mbf{x}} = D\widetilde{\theta}\,d\mbf{x}\ .
\end{equation}
Using the contour integration above, there is a residue theorem for super contour integration. If $f$ is regular at $\mbf{x}_0$, one has
\begin{equation}\label{eq:super_residue_thm}
\cint{C(\mbf{x}_0)}\! d\mbf{x}_1\, \frac{\theta_{10}}{\mbf{x}_{10}}\,f(\mbf{x}_1) = f(\mbf{x}_0)\ ,
\end{equation}
where we have absorbed a factor of $\frac{1}{2\pi i}$ into the definition of $\oint$. Furthermore, one also finds
\begin{equation}
\cint{C(\mbf{x}_0)}\! d\mbf{x}_1\, \frac{1}{\mbf{x}_{10}}\,f(\mbf{x}_1) = Df(\mbf{x}_0)\ .
\end{equation}

\subsection{Superconformal field theory}\label{ssec:SCFT}

The superconformal structure of a SRS allows for the geometric definition of a superconformal field theory (SCFT) \cite{Friedan:1986rx}.

The basic objects of a SCFT are super primaries. A chiral super primary of dimension $h$ is a field $\Phi(\mbf{x}) = \varphi(x)+\theta\,\psi(x)$, which transforms as
\begin{equation}\label{eq:super_primary}
(D\widetilde{\theta})^{2h}\,\widetilde{\Phi}(\widetilde{\mbf{x}}) = \Phi(\mbf{x})
\end{equation}
under superconformal transformations, meaning that $\Phi(\mbf{x})\,d\mbf{x}^{2h}$ is coordinate invariant. Both components of $\Phi$ are usual primaries; specialising to a superconformal coordinate transformation of the form $\widetilde{x} = g(x)$, $\widetilde{\theta} = \theta\,\sqrt{\partial g}$, one sees that $\varphi$ is a conformal primary of dimension $h$, while $\psi$ has dimension $h+1/2$.

Suppose we are given a CFT with $\mcl{N}=1$ supersymmetry, energy momentum tensor $L$, and supercharge $G$. In the NS sector, for a state $\ket{\phi}$ of dimension $h$, there is an associated super vertex operator
\begin{equation}
\Phi(\mbf{x})=\mbf{V}(\ket{\phi},\mbf{x}) := V(\ket{\phi},x) + \theta\,V(G_{-1/2}\ket{\phi},x)\ ,
\end{equation}
where $V(\ket{\phi},x)$ is the usual vertex operator. If the state $\ket{\phi}$ is annihilated by all the positive modes of $L$ and $G$, this super vertex operator is a super primary; the components $\varphi,\psi$ of $\Phi$ are thus superpartners, combined into a single field.

The global modes $L_{\pm 1},L_0,G_{\pm 1/2}$ form an $\mfr{osp}(1|2)$ algebra and implement the transformation of quasi superprimary vertex operators under super M\"obius transformations. In particular, the modes $G_{\pm 1/2}$ are the generators of supertranslations and superscalings in eqs.~(\ref{eq:super_translation}) and (\ref{eq:super_scaling}). Explicitly,
\begin{align}
e^{\lambda\,G_{-1/2}}\,\mbf{V}(\ket{\phi},x,\theta)\,e^{-\lambda\,G_{-1/2}} &= (1)^{2h}\,\mbf{V}(\ket{\phi},x-\lambda\,\theta,\theta+\lambda)\ ,\\
e^{\kappa\,G_{1/2}}\,\mbf{V}(\ket{\phi},x,\theta) \,e^{-\kappa\,G_{1/2}} &= (1-\kappa\,\theta)^{2h}\,\mbf{V}(\ket{\phi},x-\kappa\,\theta\,x,\theta + \kappa\,x)\ ,
\end{align}
as one can see by applying both sides to the vacuum \cite{Goddard:1989dp}.\footnote{Note that because $G$ is nilpotent, the exponential $e^{\lambda\,G_{-1/2}}$ is simply given by $1+\lambda\,G_{-1/2}$.} The terms $(\cdot)^{2h}$ are precisely the superconformal factors $\bbr{D\tilde{\theta}}^{2h}$ for the corresponding transformations, as in eq.~(\ref{eq:super_primary}).

The OPE of two super vertex operators has a nice form in superconformal coordinates. If we write $\Phi(\mbf{x}) = \varphi(x)+\theta\,\psi(x)$, then
\begin{equation}\label{eq:superOPE}
\Phi(\mbf{x}_1)\,\mbf{V}(\ket{\chi},\mbf{x}_0) = \sum_{r\in\mbb{Z}+h}\frac{1}{\mbf{x}_{10}^{r+h}}\,\big(\mbf{V}(\varphi_r\ket{\chi},\mbf{x}_0)+\theta_{10}\,\mbf{V}(\psi_{r-1/2}\ket{\chi},\mbf{x}_0)\big)\ .
\end{equation}
This follows by expanding both fields in components and using the usual CFT OPE \cite{Goddard:1989dp}, as well as \cite{Friedan:1986rx}
\begin{equation}
[\theta G_{-1/2},\varphi_r]=\theta\psi_{r-1/2}\ ,\quad [\theta G_{-1/2},\psi_m] = -((h-1/2)+m)\,\theta\varphi_{m-1/2}\ .
\end{equation}
Note that the OPE depends only the superdifferences and is thus manifestly supertranslation invariant. The action of the modes of the component fields can be extracted using contour integration, see eq.~(\ref{eq:super_residue_thm}),
\begin{subequations}\label{eq:super_mode_expansion}
\begin{align}
\underset{\mbf{C}(\mbf{x}_0)}{\oint}\!\! d\mbf{x}_1\,\,\theta_{10}\,\mbf{x}_{10}^{r+h-1}\,\Phi(\mbf{x}_1)\,\mbf{V}(\ket{\chi},\mbf{x}_0) &= \mbf{V}(\varphi_r\ket{\chi},\mbf{x}_0)\ ,\\
\underset{\mbf{C}(\mbf{x}_0)}{\oint}\!\! d\mbf{x}_1\,\,\mbf{x}_{10}^{m+h-1/2}\,\Phi(\mbf{x}_1)\,\mbf{V}(\ket{\chi},\mbf{x}_0) &= \,\mbf{V}(\psi_m\ket{\chi},\mbf{x}_0)\ .
\end{align}
\end{subequations}
We will use this to argue in the Section \ref{sec:orbifold} that our proposed form for the super covering map is sensible in the symmetric product orbifold description.

\section{Super covering maps}\label{sec:super_covering_maps}

We now describe super covering maps, which are the main objects of this paper. They are maps between two super Riemann surfaces\footnote{We restrict ourselves to minimal supersymmetry. One can also define $\mcl{N}=2,4$ SRSs, see for example \cite{Dorrzapf:1997rx}, and one could define super covering maps also in this context, for example by generalising the local characterisation explained in Appendix~\ref{app:structure}. However, as we can always choose an $\mcl{N}=1$ substructure, this case will always be relevant.} satisfying superconformality and regularity properties.

We will define these maps and give some simple examples.

\subsection{Definition}\label{ssec:definition}

Consider two SRS $\Sigma_1, \Sigma_2$, with local coordinates $\mbf{z} = \bbr{z,\Theta}$ and $\mbf{x} = \bbr{x,\theta}$, respectively. We want to describe specific maps
\begin{equation}
\bm{\Gamma}:\,\Sigma_1 \longrightarrow \Sigma_2;\,\,\mbf{z} \longmapsto \bbr{x,\theta} = \bbr{\Gamma_1(\mbf{z}),\Gamma_2(\mbf{z})}\ ,
\end{equation}
which behave well in the context of a SCFT.

First, such a map should respect the superconformal structure $\mcl{D}_i\subseteq T\Sigma_i$, i.e.~
\begin{equation}
\mrm{d}\bm{\Gamma}\bbr{\mcl{D}_1} \subseteq \mcl{D}_2\ ,
\end{equation}
where $\mrm{d}\bm{\Gamma}$ denotes the differential of $\bm{\Gamma}$.
In local coordinates, this can be written as
\begin{verticallines}[black]
\begin{equation}\label{eq:gamma_sc}
D\Gamma_1 = \Gamma_2\,D\Gamma_2\ ,\qquad D = \frac{\partial}{\partial \Theta} + \Theta\,\frac{\partial}{\partial z} \ .
\end{equation}
\end{verticallines}

We discussed the ramification structure of ordinary covering maps in Section \ref{ssec:cov_maps_in_adscft}. Super covering maps should have a similar structure of ramification- and branch points. The correct behaviour can be motivated as follows. Each superconformal $\bm{\Gamma}$ gives rise to an ordinary branched covering between the reduced manifolds,
\begin{equation}
\Gamma^\text{red}:\,\,\Sigma_1^\text{red} \longrightarrow \Sigma_2^\text{red}\ .
\end{equation}
If there are no odd moduli, we should recover the non-supersymmetric case $\Gamma_1 = \Gamma^\text{red}$. If $z_i$ is a ramification point of $\Gamma^\text{red}$, the behaviour of $\Gamma_1$ near $\mbf{z}_i = (z_i,0)$ is
\begin{equation}\label{eq:bos_reg}
\Gamma_1\bbr{z,\Theta} - x_i = a_i\,\bbr{z-z_i}^w + \dots\ .
\end{equation}
The branch point is $\mbf{x}_i = (x_i,0)$. The local behaviour for arbitrary ramification points $\mbf{z}_i = \bbr{z_i,\Theta_i}$ and branch points $\mbf{x}_i = \bbr{x_i,\theta_i}$ can be found by supertranslating
\begin{equation}
(z,\Theta)\mapsto (z-\Theta\,\Theta_i,\Theta+\Theta_i)\quad \text{and}\quad (x,\theta)\mapsto (x-\theta\,\theta_i,\theta+\theta_i) \ .
\end{equation}
The above regularity condition is then transformed to
\begin{equation}
\Gamma_1(\mbf{z})-x_i-\Gamma_2(\mbf{z})\,\theta_i = a_i\,(z-z_i-\Theta\,\Theta_i)^{w_i}+\mcl{O}\big((z-z_i-\Theta\,\Theta_i)^{w_i+1}\big)\ ,\quad \mbf{z}\to\mbf{z}_i\ .
\end{equation}
We should note that this is not the only way one could extend the reduced regularity in eq.~(\ref{eq:bos_reg}), as one can add nilpotent terms with lower power (see the example in eq.~(\ref{eq:non_example})). However, we want the series to start only at $(z-z_i-\Theta\,\Theta_i)^{w_i}$, as the map should lift $w_i$ twisted fields; we will discuss this further below, as well as in Appendix~\ref{app:structure}. Thus, near a ramification point $\mbf{z}_i$ we impose the regularity condition\footnote{We introduced the subscript $0$ for the free variable $\mbf{z}_0$ to be able to make use of the short-hand $\mbf{z}_{0i}$.}
\begin{verticallines}[black]
\begin{equation}\label{eq:gamma_reg}
\Gamma_1(\mbf{z}_0)-x_i-\Gamma_2(\mbf{z}_0)\,\theta_i=a_i\,\mbf{z}_{0i}^{w_i}+\mcl{O}\big(\Theta_{0i}\,\mbf{z}_{0i}^{w_i}\big)\ ,\quad \mbf{z}_0\to\mbf{z}_i\ .
\end{equation}
\end{verticallines}
Note that $a_i$ is not in general the same as in the bosonic case, but can depend on the odd moduli $\theta_i,\Theta_i$. However, $a_i$ is not allowed to be nilpotent, as we discuss in Appendix \ref{app:structure}.

In summary, a \textit{super covering map} is a map between SRSs satisfying the superconformality condition eq.~(\ref{eq:gamma_sc}). Furthermore, it has isolated ramification points $\mbf{z}_i$ with ramification index $w_i$, where it satisfies the regularity condition eq.~(\ref{eq:gamma_reg}).

\subsection{Properties}

We briefly discuss some of the properties of the super covering maps defined above.

Near a ramification point, we can also describe what the second component of $\bm{\Gamma}$, $\Gamma_2$, looks like. Taking the superderivative of eq.~(\ref{eq:gamma_reg}) and demanding superconformality (\ref{eq:gamma_sc}), one finds that around $\mbf{z}_i$, $\Gamma_2$ behaves as
\begin{equation}\label{eq:g2_reg}
\Gamma_2(\mbf{z}_0)-\theta_i = \sqrt{w_i\,a_i}\,\Theta_{0i}\,\mbf{z}_{0i}^{\frac{w_i-1}{2}}+\mcl{O}\big(\mbf{z}_{0i}^{\frac{w_i+1}{2}}\big)\ ,\quad \mbf{z}_0\to \mbf{z}_i\ .
\end{equation}
Around an insertion of a twist operator of cycle length $w_i$, we see that $w_i$ must be odd in order for $\mbf{z}_{0i}^{\frac{w_i-1}{2}}$ to be single valued. This is sensible, as we can see from the non-supersymmetric analysis \cite{Lunin:2001pw}; for even $w_i$, the fermions on the covering surface are in the Ramond sector. Thus, the correct notion of a puncture in this case is that of an R puncture, as we discuss below in Section \ref{ssec:R_punctures}.

The local structure of a super covering map is also interesting. The superconformality in eq.~(\ref{eq:gamma_sc}) implies that, away from the ramification points, $\bm{\Gamma}$ is locally a change of coordinates. Furthermore, one can always find coordinates in which $\bm{\Gamma}$ is simple, as we discuss in Appendix~\ref{sapp:local_expression}. Around a point of ramification index $w$, one can pass to local charts on $\Sigma_1,\Sigma_2$, such that
\begin{equation}
\Gamma_1\bbr{z,\Theta} = z^w\ ,\qquad \Gamma_2\bbr{z,\Theta} = \sqrt{w}\,\Theta\,z^{\frac{w-1}{2}}\ .
\end{equation}

\subsection{Genus zero}\label{ssec:genus_zero}

To give examples, let us consider super coverings of the supersphere $\mbb{CP}^{1|1}$ by itself. We want to describe the form of $\bm{\Gamma}:\,\mbb{CP}^{1|1}\longrightarrow \mbb{CP}^{1|1}$ for the situation of $n$ NS punctures with (odd) ramifications $w_i$. The ordinary covering map $\Gamma^\mrm{red}$ underlying $\bm{\Gamma}$ is a rational function of degree
\begin{equation}
N = 1+\sum_{i=1}^n\frac{w_i-1}{2}
\end{equation}
by the Riemann-Hurwitz formula. One can show that $\bm{\Gamma} = \big(\Gamma_1,\Gamma_2\big)$ is of the form\footnote{The non-trivial statement here is the truncation of the nilpotent terms. This follows from the global properties of $\bm{\Gamma}$, more specifically the regularity under the inversion of $\mbb{CP}^{1|1}$.}
\begin{align}\label{eq:rational_form}
\Gamma_1(\mbf{z}) &:= \frac{P(\mbf{z})}{Q(\mbf{z})}= \frac{\big(\sum_{k=0}^{N-1} c_k\,z^k+\alpha_k\,\Theta\,z^k\big) +c_N\,z^N}{\big(\sum_{k=0}^{N-1} d_k\,z^k+\beta_k\,\Theta\,z^k\big) +d_N\,z^N}\ ,\\
\Gamma_2(\mbf{z}) &:= \frac{\Upsilon(\mbf{z})}{Q(\mbf{z})}=\frac{\big(\sum_{k=0}^{N-1} \epsilon_k\,z^k+r_k\,\Theta\,z^k\big) +\epsilon_N\,z^N}{\big(\sum_{k=0}^{N-1} d_k\,z^k+\beta_k\,\Theta\,z^k\big) +d_N\,z^N}\ ,\nonumber
\end{align}
where greek letters are odd and latin letters are even variables.

The superconformality condition eq.~(\ref{eq:gamma_sc}) can be written as
\begin{equation}\label{eq:poly_sc}
DP\,Q - P\,DQ = \Upsilon\,D\Upsilon\ .
\end{equation}
Furthermore, a ramification point $\mbf{z}_i$ is characterised by the regularity condition
\begin{align}\label{eq:poly_reg}
P(\mbf{z}_0)-Q(\mbf{z}_0)\,x_i-\Upsilon(\mbf{z}_0)\,\theta_i&=\mcl{O}\big(\mbf{z}_{0i}^{w_i})\ ,\\
\Upsilon(\mbf{z}_0)-Q(\mbf{z}_0)\,\theta_i&=\mcl{O}\big(\Theta_{0i}\,\mbf{z}_{0i}^{(w_i-1)/2}\big)\ ,\quad i=1,\dots,n\ .\nonumber
\end{align}
These equations only have solutions on isolated points in the moduli space of the punctured $\mbb{CP}^{1|1}$, as one can see by counting the parameters. The odd function $\Upsilon$ is fixed by eq.~(\ref{eq:poly_sc}), and so the free parameters are those of $P,Q$. There are $2N+1$ even parameters (as there is an overall irrelevant scale), and $2N$ odd parameters. The first line in eq.~(\ref{eq:poly_reg}) imposes $w_i$ even and odd constraints for each ramification point. In total, there are thus $2N+n-2$ even and odd constraints. The system is therefore overdetermined, with codimension $n-3|n-2$, which is just the dimension of the moduli space of the punctured supersphere.

Let us show some examples of super covering maps with three ramification points $\mbf{z}_i$ with ramification indices $w_i$. Using the $3|2$ dimensional super M\"obius transformations, we can always fix, see the discussion around eq.~(\ref{eq:reference_points}),
\begin{equation}
\mbf{z}_1 = \bbr{\infty,0}\ ,\quad \mbf{z}_2 = \bbr{1,0}\ ,\quad \mbf{z}_3 = \bbr{0,\Theta_3}\ ,
\end{equation}
as well as
\begin{equation}
\mbf{x}_1 = \bbr{\infty,0}\ ,\quad \mbf{x}_2 = \bbr{1,0}\ ,\quad \mbf{x}_3 = \bbr{0,\theta_3}\ ,
\end{equation}
for the branch points $\mbf{x}_i = \bm{\Gamma}(\mbf{z}_i)$. In practice, the covering map can be found by writing the general rational function in eq.~(\ref{eq:rational_form}) and imposing the regularity conditions, eq.~(\ref{eq:poly_reg}). Any remaining freedom can then be fixed by imposing superconformality, eq.~(\ref{eq:poly_sc}), which is a non-linear relation.

\begin{verticallines}[black!30!white]
%\hspace{-\parindent}
\textbf{Example 1.} For the ramification profile $(w_1,w_2,w_3)=(5,3,3)$, the covering map is polynomial and takes the form
\begin{align}\label{eq:cov_533}
\Gamma_1(\mbf{z}) &= z^3\bbr{6z^2-15 z+10} + \sqrt{30}\,z\,\bbr{z-1}^3\,\Theta\,\theta_3\ ,\\
\Gamma_2(\mbf{z}) &= \bbr{z-1}\Big(z\,\bbr{\sqrt{30}\,\Theta + \theta_3} - \theta_3\Big) \ .\nonumber
\end{align}
The parameter $\Theta_3$ of the ramification point $\mbf{z}_3=\bbr{0, \Theta_3}$ is fixed by the equations and equals
\begin{equation}
\Theta_3 = -\sqrt{\frac{2}{15}}\,\theta_3\ .
\end{equation}
For $\theta_3=0$, one recovers the bosonic covering map
\begin{equation}
\Gamma_1(\mbf{z})\Big|_{\theta_3=0} = \Gamma^\text{red}(z) \ ,\qquad \Gamma_2(\mbf{z})\Big|_{\theta_3=0} = \Theta\,\sqrt{\partial\Gamma^\text{red}}(z) \ .
\end{equation}
\end{verticallines}

\begin{verticallines}[black!30!white]
%\hspace{-\parindent}
\textbf{Example 2.} For the ramification profile $(3,3,3)$, the covering map is a rational function of degree 4,
\begin{align}\label{eq:cov_333}
\Gamma_1(\mbf{z}) &= -\frac{z^3\bbr{z-2}+i\sqrt{\frac{2}{3}}\,z\,\bbr{z^2-3\,z+3}\,\Theta\,\theta_3}{2\,z -1 - i\sqrt{\frac{2}{3}}\,\Theta\,\theta_3}\ ,\\
\Gamma_2(\mbf{z}) &= \sqrt{\frac{3}{2}}\,\,\frac{\bbr{z-1}\Big(-2i\,z\,\Theta + \sqrt{\frac{2}{3}}\bbr{z-1}\theta_3\,\Big)}{2\,z -1 - i\sqrt{\frac{2}{3}}\,\Theta\,\theta_3} \ .\nonumber
\end{align}
The modulus $\Theta_3$ is again fixed by the equations,
\begin{equation}
\Theta_3 = 0\ .
\end{equation}
The ramification point on the covering surface thus never has an odd component, regardless of the choice of $\theta_3$. This can be explained by a Ward identity in the symmetric orbifold, implying the vanishing of certain correlators for this ramification profile, as will be discussed in Section \ref{ssec:orbifold_examples}.
\end{verticallines}

For covering maps with three ramification points there is only one odd modulus, and thus the (Grassmann even) expansion coefficients $a_i$ around the $\mbf{z}_i$ equal those of the underlying map $\Gamma^\text{red}$. This is no longer the case for maps with four or more ramification points.

\begin{verticallines}[black!30!white]
%\hspace{-\parindent}
\textbf{Example 3.} The covering maps for the ramification profile $(3,3,3,3)$ is a rational function of degree 5. As the even parts of the ramification points are no longer entirely fixed by the M\"obius transformations, there are many possible covering maps. Let us focus on the supersymmetric extension of the bosonic covering map
\begin{equation}
\Gamma^\text{red}(z) = -z^3\frac{z^2-5\,z + 5}{5\,z^2 - 10\,z+4} \ ,
\end{equation}
which has ramification and branch points
\begin{equation}
z_1=x_1=\infty\ ,\quad z_2=x_2=2\ ,\quad z_3=x_3=1\ ,\quad z_4=x_4=0\ .
\end{equation}
The super covering map $\bm{\Gamma}$ with branch points
\begin{equation}
\mbf{x}_1 = \bbr{\infty,0}\ ,\quad \mbf{x}_2 = \bbr{2,0}\ ,\quad \mbf{x}_3 = \bbr{1,\theta_3}\ ,\quad \mbf{x}_4 = \bbr{0,\theta_4}\ ,
\end{equation}
can be found using the procedure outlined above. We do not write it down here, but rather state its ramification points and Taylor coefficients. The map $\bm{\Gamma}$ is ramified at
\begin{equation}
\mbf{z}_1 = \bbr{\infty,0}\ ,\  \mbf{z}_2 = \bbr{2,0}\ ,\  \mbf{z}_3 = \bbr{1,-\tfrac{i}{\sqrt{15}}(\theta_3-\theta_4)}\ ,\  \mbf{z}_4 = \bbr{\tfrac{8}{15}\theta_3\theta_4, -\tfrac{i}{\sqrt{15}} (2\theta_3-\theta_4) }\ .
\end{equation}
The leading coefficients in the Taylor series at the ramification points are
\begin{align}\label{eq:cov_3333}
a_1 &= -\frac{1}{5}\,\bbr{1+\tfrac{3}{5}\theta_3\theta_4}\ , &\quad a_2 &= -\frac{1}{4}\,\bbr{5+\theta_3\theta_4}\ ,\\
a_3 &= -5 + \theta_3\theta_4\ , &\quad a_4 &= -\frac{1}{4}\bbr{5+3\theta_3\theta_4}\ .\nonumber
\end{align}
The $a_i$ are thus modified by nilpotent terms. We can also extract the poles $\bm{\zeta}_j=\bbr{\zeta_j,\varphi_j}$, where
\begin{equation}
\Gamma_1(\mbf{z}) = \frac{C_j}{z-\zeta_j - \Theta\,\varphi_j} + \dots\ .
\end{equation}
One finds two poles
\begin{equation}
\bm{\zeta}_\pm = \Big(1\pm\tfrac{1}{\sqrt{5}} +\big(-1\pm \tfrac{3}{\sqrt{5}}\big)\Theta_3\Theta_4, 2\Theta_3-\tfrac{1}{2}\big(1\pm \sqrt{5}\big)\Theta_4\Big) \ ,
\end{equation}
where $\Theta_i$ is the odd component of $\mbf{z}_i$ given above.
The pole coefficients at these points are given by
\begin{equation}
C_\pm = \frac{12}{125} - \frac{4}{625}\bbr{8-5\sqrt{5}}\theta_3\theta_4\ .
\end{equation}
These terms correctly encode the supersymmetric structure of the SCFT, as we will discuss in Section \ref{ssec:twist_correlator}.
\end{verticallines}

\subsection{Higher genus}

When one considers super covering maps between super Riemann surfaces of higher genus, new effects due to the complicated structure of the super moduli spaces \cite{Witten:2012ga} have to be taken into account. We briefly describe some of these features and illustrate them with a super torus covering of the super sphere.

Let us first discuss the case with no odd moduli, where we see that a compatibility issue of spin structures arises in the definition of a super covering map. We consider a super covering map
\begin{equation}
\bm{\Gamma}:\,\Sigma_1 \longrightarrow \Sigma_2
\end{equation}
where $\Sigma_2$ is split and the branch points $\mbf{x}_i = \bbr{x_i,0}\in \Sigma_2$ have no odd components. As a consequence, the covering surface $\Sigma_1$ must also be split and the ramification points are $\mbf{z}_i=\bbr{z_i,0}$.\footnote{The odd moduli on the covering surface must be functions of the base space moduli. Thus, if all the $\theta_i=0$ and $\Sigma_2$ is split, all the odd moduli are zero.} Locally, then, $\bm{\Gamma}$ is determined by an ordinary covering map $\Gamma^\text{red}:\Sigma_1^\text{red}\to\Sigma_2^\text{red}$
\begin{equation}
\Gamma_1\bbr{\mbf{z}} = \Gamma^\text{red}(z)\ ,\qquad \Gamma_2\bbr{\mbf{z}} = \Theta\,\sqrt{\partial \Gamma^\text{red}}(z)\ .
\end{equation}
While $\Gamma^\text{red}$ exists independently of additional structures on the Riemann surface, $\bm{\Gamma}$ does not. As explained in \cite[Section~2]{Witten:2012ga}, the odd coordinates $\Theta$, $\theta$ transform as fermionic sections of a square root of the cotangent bundle on $\Sigma_1^\text{red}$, $\Sigma_2^\text{red}$, respectively, and hence a spin structure is chosen in the definition of a SRS. The split $\Sigma_i$ can be seen as a dual bundle of the square root of the cotangent bundle over $\Sigma_i^\text{red}$ with fermionic fibres. The superconformality condition of $\bm{\Gamma}$ means that the fibres of $\Sigma_1$ are mapped onto those of $\Sigma_2$ by the pushforward of $\Gamma^\text{red}$. The identification $\theta = \Theta\,\sqrt{\partial \Gamma^\text{red}}$ thus imposes a pullback condition on the spin structure on $\Sigma_1^\text{red}$, as we will see in the following example.

\begin{verticallines}[black!30!white]
%\hspace{-\parindent}
\textbf{Example.} To see the matching of spin structures in an explicit instance, consider the following torus covering of the sphere:
\begin{equation}\label{eq:torus_covering}
\Gamma^\text{red}:\,\,\mbb{T}_{\tau_*} \longrightarrow \mbb{CP}^1;\quad z \longmapsto \frac{1}{2} - \frac{\wp'\bbr{z;\tau_*}}{2\,\wp'\bbr{a;\tau_*}}\ ,
\end{equation}
where
\begin{equation}
a = \frac{\sqrt{3}+i}{2\sqrt{3}}\ ,\quad \tau_* = \frac{1+\sqrt{3} i}{2}\ ,
\end{equation}
and $\wp(z;\tau)$ is the Weierstrass elliptic function. This covering map has three ramification points of ramification index 3 at $z_1 = 0$, $z_2 = 2a$, $z_3 = a$, with branch points $x_1=\infty$, $x_2=1$, $x_3=0$.\\
Let us investigate how this can be extended to a super covering map $\bm{\Gamma}$ from a super torus $\mbb{T}$ to $\mbb{CP}^{1|1}$. For the simplest case with no odd modulus $\theta_3=0$, the components of the super covering map are
\begin{equation}
\Gamma_1\bbr{z,\Theta} = \Gamma^\text{red}(z)\ ,\quad \Gamma_2\bbr{z,\Theta} = \Theta\,\sqrt{\partial \Gamma^\text{red}}(z) = i\sqrt{\frac{3}{\wp'\bbr{a;\tau_*}}}\,\Theta\,\wp\bbr{z;\tau_*}\ .
\end{equation}
Note that $\Gamma_2$ is elliptic; this implies that the underlying covering map can only be extended to the (split) super torus with odd spin structure. For an easy way to see this, take a non-contractible cycle $\gamma:[0,1]\to \mbb{T}_{\tau_*}$. The spin structure determines a sign in the identification on $\mbb{T}_{\tau_*}$ \cite{Rabin:1987rg}
\begin{equation}
\bbr{\gamma(0),\Theta} \sim \bbr{\gamma(1), \pm \Theta}\ .
\end{equation}
However, on the super torus covering the super sphere, only the `+' sign can arise, since
\begin{equation}
\bbr{x,\theta} = \bm{\Gamma}\bbr{\gamma(0),\Theta} = \bm{\Gamma}\bbr{\gamma(1),\pm \Theta} = \bbr{x,\pm \theta}\ ,
\end{equation}
as the spin structure on the sphere is unique. Thus, the spin structure on $\mbb{T}_{\tau_*}$ is the trivial bundle, i.e.~the odd spin structure.
\end{verticallines}

When $\Sigma_2$ is deformed by odd moduli, i.e.~when the SRS is no longer split or the odd coordinates of the ramification points $\mbf{x}_i=\bbr{x_i,\theta_i}$ are not zero, the covering SRS $\Sigma_1$ is also deformed. In general, the ramification points $\mbf{z}_i=\bbr{z_i,\Theta_i}$ can have odd components; these moduli behave similarly for genus zero and higher genus coverings. More interesting are the odd moduli which deform the unpunctured $\Sigma_1$ away from the split point. On unsplit SRSs, the even and odd coordinates can no longer be disentangled, but will always mix in some transition functions, making them more intricate than ordinary Riemann surfaces with spin structures. Importantly, the odd moduli take values in a space which depends on the spin structure of the underlying Riemann surface \cite{Witten:2012ga}. If one wants to find a super covering map extending an underlying $\Gamma^\text{red}$, one thus has to first find the spin structure and the appropriate space of odd moduli by studying the split case, as discussed above. The values the odd moduli take can then in principle be found by solving the superconformality and regularity conditions given in eqs.~(\ref{eq:gamma_sc}) and (\ref{eq:gamma_reg}).\\
Below, we show an example of a super covering map from an unsplit super torus to the super sphere, which illustrates some interesting features of such coverings. Firstly, it shows that the covering surface of a split SRS is not necessarily split itself; the odd coordinates $\theta_i$ of the branch points can deform the covering surface away from the split point. Secondly, the functional form of $\bm{\Gamma}$ can be qualitatively different from the underlying $\Gamma^\text{red}$. In particular, we will see that the $z$ dependence of $\bm{\Gamma}$ need not be elliptic, because the mixing of even and odd coordinates necessitates the consideration of the non-elliptic function $\dot\wp(z,\tau):=\partial_\tau \wp(z,\tau)$ \cite{Rabin:1987rg}. This is in contrast to the genus zero case, where the $z$ dependence is always rational. For higher genus SRSs, super covering maps are thus somewhat more complicated objects than the underlying ordinary covering maps.

\begin{verticallines}[black!30!white]
%\hspace{-\parindent}
\textbf{Example (continued).}
We now consider the torus covering from the previous example in the most general case of branch points $\mbf{x}_1=\bbr{\infty,0}$, $\mbf{x}_2=\bbr{1,0}$, $\mbf{x}_3=\bbr{0,\theta_3}$. The analysis of the split case tells us that when we turn on the $\theta_3$ modulus on the punctured $\mbb{CP}^{1|1}$, we vary the covering surface in the super moduli space of super tori with \emph{odd} spin structure.\\
Let us be specific about how this can work in practice. The relevant super tori are described as a quotient of $\mbb{C}^{1|1}$ by the group generated by the supertranslations \cite{Rabin:1993bw}
\begin{equation}
\bbr{z,\Theta} \longmapsto \bbr{z+1,\Theta}\ ,\qquad \bbr{z,\Theta} \longmapsto \bbr{z+\tau + \Theta\, \delta, \Theta+\delta}\ ,
\end{equation}
where $\delta$ is an odd modulus. Maps from this torus were analysed in \cite{Rabin:1987rg}, where it was shown that every super elliptic map can be written in terms of the super Weierstrass function
\begin{equation}
R(z,\Theta) = \wp(z;\tau+\Theta \delta)=\wp(z;\tau) + \Theta\,\delta\,\dot\wp(z;\tau)\ ,\qquad \dot\wp=\partial_\tau\wp\ ,
\end{equation}
and its derivatives. Making the most general ansatz for $\bm{\Gamma}$ extending $\Gamma^\text{red}$ in eq.~(\ref{eq:torus_covering}) and imposing the regularity conditions eq.~(\ref{eq:gamma_reg}) fixes the super covering map
\begin{align}
\Gamma_1\bbr{z,\Theta} &= \frac{1}{2} + c^2\,\dot{g}_2(\tau_*)\,\Theta\,\delta\,\wp -\sqrt{3}i\,c \,\Theta\,\delta\,\wp' + c\,\bbr{\wp' + \Theta\,\delta\,\dot\wp'}\ ,\nonumber\\
\Gamma_2\bbr{z,\Theta} &= \sqrt{\frac{c}{6}}\,\left( 6\,\Theta\,\wp + \delta\,\frac{\dot\wp'-\sqrt{3}i\,\wp'}{\wp}+\delta\,c\,\dot{g}_2(\tau_*) \right)\ ,
\end{align}
where $c=- \bbr{2\,\wp'\bbr{a;\tau_*}}^{-1}$, and $\dot{g}_2$ is the $\tau$ derivative of the Weierstrass invariant. The modulus $\delta$ is fixed by requiring $\theta_3 = \Gamma_2\bbr{a,\Theta_3}$, which gives
\begin{equation}
\theta_3 = \sqrt{\frac{2\,c^3}{3}}\,\dot{g}_2(\tau_*)\,\delta\ .
\end{equation}
In this way, the conditions imposed on the super covering map define a point in the super moduli space of covering tori.

\end{verticallines}

\subsection{Ramond punctures}\label{ssec:R_punctures}

Let us also discuss super covering maps in the case where the covering SRS has Ramond punctures. As we saw in eq.~(\ref{eq:g2_reg}), the odd component of $\bm{\Gamma}$ looks multivalued around a NS puncture with even ramification $w$. As the odd coordinate $\theta$ on the base space is single valued when encircling the branch point, this implies that the odd coordinate $\Theta$ on the covering surface must be multi-valued. The reason is clear from the perspective of the symmetric orbifold: if one inserts a twist operator with even cycle length, it is lifted to a spin operator on the covering surface, meaning that the fermions are in the Ramond sector \cite{Lunin:2001pw}. The proper notion of puncture to use in the case of even ramification indices is thus that of a Ramond puncture, which is discussed in \cite[Section~4.1.3]{Witten:2012ga}. There are two descriptions of such a Ramond puncture: either $\Theta$ is multi-valued and ill-defined at the puncture, or one replaces $\Theta$ by a single-valued $\tilde{\Theta}$ at the cost of changing the superconformal structure. The former description is more suitable for describing vertex operators in a SCFT, while the global structure of the SRS is more apparent in the latter description. Let us explain this a bit further.

In the global description, a Ramond puncture is not a point but rather a $0|1$ dimensional submanifold of the SRS \cite{Witten:2012ga}. At a R puncture, the superconformal structure $\mcl{D}$ degenerates, and one can always find superconformal coordinates $\bbr{z,\tilde{\Theta}}$ around the R puncture in which the superderivative becomes
\begin{equation}
D = \frac{\partial}{\partial \tilde{\Theta}} + z\,\tilde{\Theta}\,\frac{\partial}{\partial z}\ .
\end{equation}
In these coordinates, the R puncture is the submanifold $\big\{\bbr{z=0,\tilde{\Theta}}\big\}$.
The conditions for $\bm{\Gamma}$ in the presence of R punctures are now formally unchanged, see eqs.~(\ref{eq:gamma_sc}), (\ref{eq:gamma_reg}), but the superderivative $D$ is modified. Concretely, if the even part of $\bm{\Gamma}$ behaves as
\begin{equation}
\Gamma_1\bbr{z,\tilde{\Theta}} = a\,z^w + \dots\ ,\qquad w \in 2\mbb{N}\ ,
\end{equation}
near the R puncture, the superconformality condition fixes
\begin{equation}
\Gamma_2\bbr{z,\tilde{\Theta}} = \sqrt{w a}\,\tilde{\Theta}\,z^{\frac{w}{2}} + \dots\ .
\end{equation}
Thus, $\Gamma_2$ is single valued in these coordinates, and is globally defined without branch cuts.

To pass to the second description of the R puncture, one can define
\begin{equation}
\Theta = \sqrt{z}\,\tilde{\Theta}\ .
\end{equation}
This coordinate is now multi-valued around $z=0$, and one recovers the behaviour
\begin{equation}
\Gamma_2\bbr{z,{\Theta}} = \sqrt{w a}\,{\Theta}\,z^{\frac{w-1}{2}} + \dots
\end{equation}
for the odd component, as in eq.~(\ref{eq:g2_reg}). An important point is that $\Gamma_2$ is still single-valued, because the monodromies of $\Theta$ and $z^{\frac{w-1}{2}}$ around $z=0$ cancel.

The upshot of this discussion is the following: if we want to see the global structure super covering map, and in particular its single-valuedness, explicitly, it is better to use the $\bbr{z,\tilde{\Theta}}$ coordinates. On the other hand, if we work with fields in a SCFT, it is more natural to use the multi-valued $\Theta$, as the fermionic fields are also multi-valued in the presence of a spin field. In particular, in the context of the symmetric orbifold, $\Theta$ is the correct coordinate for lifting fields to the covering surface. We give an explicit example of a covering map in the presence of R punctures and discuss the relation to the orbifold in Appendix~\ref{app:R_cov_map}.

\section{Coverings in the Orbifold}\label{sec:orbifold}

We now discuss symmetric product orbifolds of superconformal field theories on super Riemann surfaces. We first show that the analytic structure of the super covering map defined in Section \ref{ssec:definition} is the correct one to lift fields in the presence of a twist operator. We then discuss how the analysis of \cite{Lunin:2000yv} extends to the supersymmetric case, expressing the ground state correlator in terms of super covering map data. We also discuss an extended example of three-point functions of primary operators. For notational simplicity, we restrict ourselves to the case of NS punctures and present the situation with R punctures in Appendix~\ref{app:R_cov_map}.

\subsection{Lifting fields}\label{ssec:lifting}

Here we discuss the way super fields are twisted by twist operators in the symmetric product orbifold of a SCFT. We show how a super covering map naturally resolves this twisting and lifts the fields to a single-valued field on the super covering surface. We focus on the case where the fields are inserted at NS punctures on the covering surface, i.e.~odd twist $w\in 2\,\mbb{N}-1$. We discuss the case with Ramond punctures in Appendix~\ref{app:R_cov_map}.

We sketched how a covering map undoes the twist in Section \ref{ssec:cov_maps_in_adscft}, but let us be more precise now, and first focus on the bosonic theory. Let $v^{(k)}$ denote a field in the $k$-th copy of the seed theory. In the presence of a twist operator $\sigma_w(x_1)$, these fields are mixed when they encircle the insertion $x_1$,
\begin{equation}
v^{(k)}\bbr{\gamma(1)}\,\sigma_w(x_1) = v^{(k+1)}\bbr{\gamma(0)}\,\sigma_w(x_1)\ ,\qquad k=1,\dots,w\ ,\quad w+1 \equiv 1\ ,
\end{equation}
where $\gamma(t)=x_1 + \varepsilon\,e^{2\pi i t}$ is a small circle around $x_1$. The covering map $\Gamma$ is ramified over $x_1$ and locally looks like
\begin{equation}
\Gamma(z) = x_1 + a\,\bbr{z-z_1}^{w} + \dots\ ,\quad z\to z_1\ .
\end{equation}
This means that each point in the vicinity of $x_1$ (but excluding $x_1$), has $w$ pre-images in the vicinity of $z_1$. The path $\gamma$ above is lifted only to a $\frac{2\pi}{w}$ circle segment around $z_1$ and connects two different pre-images of $x_1+\varepsilon$. By labelling these pre-images by $k$, the $v^{(k)}$ can be lifted to a single field $\mcl{V}$ on the covering surface. The identification on the level of operators is given by
\begin{equation}\label{eq:lifted_field_relation}
    v^{(k)}(\Gamma(z)) \ \longleftrightarrow \ \big(\partial \Gamma(z)\big)^{-h}\,\mcl{V}(z)
\end{equation}
for $z$ the pre-image labelled $k$. Here, $h$ is the conformal dimension of $v^{(k)}$, and `$\leftrightarrow$' indicates that we pass between the Hilbert spaces of the orbifold CFT and the CFT on $\Sigma$.

The situation is essentially unchanged for superfields but needs to be slightly adjusted to accomodate superconformal changes of coordinate as discussed in Section \ref{ssec:SCFT}. Let $\phi^{(k)}(\mbf{x}) = v^{(k)}(x)+\theta\,\rho^{(k)}(x)$ denote the superfield (dimension $h$) of the $k$'th seed theory copy in the orbifold theory, and let $\bm{\Gamma}=\bbr{\Gamma_1,\Gamma_2}$ be a super covering map. The fields are lifted to a single superfield $\Phi(\mbf{z}) = \mcl{V}(z)+\Theta\,\mcl{P}(z)$ on the covering $\Sigma$, such that locally, in the sense discussed above,
\begin{equation}
\phi^{(k)}\br{\bm{\Gamma}(\mbf{z})} \longleftrightarrow \big(D\Gamma_2(\mbf{z})\big)^{-2h}\,\Phi(\mbf{z})\ ,
\end{equation}
i.e.~that everywhere except at the insertions, $\Phi$ is just a superconformal coordinate change of one of the seed theory copies, which get mixed at the twist operator insertions.

With this identification, we can now check that the regularity conditions (\ref{eq:gamma_reg}) are well-suited to lift vertex operators on the base space to the covering surface. In particular, the fractional modes of the fields in the twisted sectors are lifted as expected. We will follow \cite[Section 2.2]{Gaberdiel:2022oeu}, which generalises straight-forwardly to theories on SRS.

In the $w$ twisted sector, with a twist operator inserted at $\mbf{x}_1$, the linear combinations
\begin{equation}
\psi^j(\mbf{x}) = \sum_{k=1}^w \omega^{j\,(k-1)}\,\phi^{(k)}(\mbf{x})\ ,\quad \omega = e^{2\pi i/w}\ ,
\end{equation}
pick up a phase $\omega^{-j}$ upon going around the insertion $\mbf{x}_1$ once. They can thus be mode expanded in components as
\begin{equation}
\psi^j(\mbf{x}) = \sum_{r\in w\mbb{Z}+j-h} v_{r/w}\,x^{-r/w-h} + \sum_{s\in w\mbb{Z}+j-(h+1/2)} \!\!\!\!\theta\,\rho_{s/w}\,x^{-s/w-(h+1/2)} \ .
\end{equation}

The vertex operator $\mbf{V}(v_{r/w}\ket{\chi},\mbf{x}_1)$ with $r\in w\mbb{Z}+j-h$ of a state in the $w$ twisted sector can then be written as
\begin{equation}
\underset{\mbf{C}(\mbf{x}_1)}{\oint} \!\! d\mbf{x}_0\,\,\theta_{01}\,\mbf{x}_{01}^{r/w+h-1}\,\psi^{j}(\mbf{x}_0)\,\mbf{V}(\ket{\chi},\mbf{x}_1)\ .
\end{equation}
Letting $\mbf{V}(\ket{X},\mbf{z}_1)$ be the lift of $\mbf{V}(\ket{\chi},\mbf{x}_1)$ to $\Sigma$, we can now use a superconformal change of variables from $\mbf{x}=\bm{\Gamma}(\mbf{z})$ to $\mbf{z}$ to find that this expression corresponds to
\begin{equation}
\underset{\mbf{C}(\mbf{z}_1)}{\oint}\!\!d\mbf{z}_0\,\,\big(D\Gamma_2(\mbf{z}_0)\big)^{1-2h}\,\big(\Gamma_2(\mbf{z}_0)-\theta_1\big)\,\big(\Gamma_1(\mbf{z}_0)-x_1-\Gamma_2(\mbf{z}_0)\,\theta_1\big)^{r/w+h-1}\,\Phi(\mbf{z}_0)\,\mbf{V}(\ket{X},\mbf{z}_1)
\end{equation}
on the covering surface. Using the regularity property (\ref{eq:gamma_reg}), we get a series of terms. The highest mode application of $\Phi$ comes from the leading $a\,\mbf{z}_{01}^{w}$ term in the super Taylor expansion. Explicitly, it is given by
\begin{equation}\label{eq:super_mode_lifting}
w^{1-h}\,a^{r/w}\,\underset{\mbf{C}(\mbf{z}_1)}{\oint}\!\!d\mbf{z}_0\,\,\Theta_{01}\,\mbf{z}_{01}^{r+h-1}\,\Phi(\mbf{z}_0)\,\mbf{V}(\ket{X},\mbf{z}_1) = w^{1-h}\,a^{r/w}\,\mbf{V}(\mcl{V}_r\ket{X},\mbf{z}_1)\ .
\end{equation}
Thus we recover exactly the usual mode lifting of \cite{Lunin:2000yv,Lunin:2001pw}. We can do the same for
\begin{equation}
\mbf{V}(\rho_{s/w}\ket{\chi},\mbf{x}_1)=\underset{\mbf{C}(\mbf{x}_1)}{\oint} \!\! d\mbf{x}_0\,\,\mbf{x}_{01}^{s/w+h-1/2}\,\psi^{j}(\mbf{x}_0)\,\mbf{V}(\ket{\chi},\mbf{x}_1)
\end{equation}
and again get that this is lifted to
\begin{equation}
w^{1/2-h}\,a^{s/w}\,\mbf{V}(\mcl{P}_s\ket{X},\mbf{z}_1)\ ,
\end{equation}
plus terms coming from the subleading coefficients in the Taylor expansion.
Thus the regularity properties (\ref{eq:gamma_reg}) and superconformality condition (\ref{eq:gamma_sc}) on $\bm{\Gamma}$ are precisely the conditions needed to extend the lifting of \cite{Lunin:2000yv} to the SRS setting. Furthermore, the superfield can be treated as a whole, without expanding it in even and odd components.

\subsection{Correlators}

Correlators in the orbifold of a SCFT can be calculated using the super covering maps. We first discuss the correlators of super twist fields $\bm{\sigma}_w$, where
\begin{equation}
\bm{\sigma}_w(\mbf{x}) = \sigma_w(x) + \theta\,G_{-\frac{1}{2}}\,\sigma_w(x)\ ,
\end{equation}
is the superfield associated to the (primary) twist operator $\sigma_w$.
These correlators are given by a super Weyl anomaly when passing to the covering surface. We then show an example of three-point functions of descendants of twist operators and relate the covering map branch points to Ward identities for the orbifold correlators.

\subsubsection{Correlator of twist fields}\label{ssec:twist_correlator}

The basic correlators in the symmetric product orbifold are correlation functions of twist operators. These correlators are geometrically calculated as the vacuum partition function on the covering surface \cite{Lunin:2000yv}. For sphere coverings, this is given by the Liouville action of the conformal anomaly and can be expressed in terms of local data of the covering map. This construction has an analogous supersymmetric formulation. A correlator of twist superfields
\begin{equation}
\corr{\bm{\sigma}_{w_1}\bbr{\mbf{x}_1}\cdots \bm{\sigma}_{w_n}\bbr{\mbf{x}_n}}\ ,\qquad w_i \in 2\,\mbb{N}-1\ ,
\end{equation}
is the sum of the partition functions over all the covering SRSs.

If the covering SRS has genus zero, the calculation of \cite{Lunin:2000yv} can be repeated with the superconformal anomaly action to obtain an expression of the correlator in terms of covering map data. We show the relevant notions and calculations in Appendix \ref{app:super_liouville}. For a sphere covering $\bm{\Gamma}:\mbb{CP}^{1|1}\to \mbb{CP}^{1|1}$ with ramification points $\mbf{z}_i$ and poles by $\bm{\zeta}_j$, the result can be expressed in terms of the leading coefficients in a super Taylor expansion around these points. Explicitly, denote by $a_i$ the coefficients in an expansion around $\mbf{z}_i$
\begin{equation}
\Gamma_1(\mbf{z}_0)-x_i-\Gamma_2(\mbf{z}_0)\,\theta_i=a_i\,\mbf{z}_{0i}^{w_i} + \dots\ ,
\end{equation}
and by $C_j$ the residues at a pole $\bm{\zeta}_j=\bbr{\zeta_j,\varphi_j}$
\begin{equation}
\Gamma_1(\mbf{z}) = \frac{C_j}{z-\zeta_j - \Theta\,\varphi_j} + \dots\ .
\end{equation}
Then, the correlator of twist fields is given by\footnote{Here, we fix one of the poles to be at $\infty$ w.l.o.g., and the corresponding residue $C_\infty$ is not included in the product. This follows the derivation in \cite{Lunin:2000yv} and will be apparent in Appendix~\ref{app:super_liouville}. If this choice is not made, there is a slight modification to the formula, as discussed in \cite{Hikida:2020kil} and \cite[Footnote~22]{Dei:2023ivl}.}
\begin{verticallines}[black]
\vspace{-0.5em}
\begin{equation}\label{eq:super_corr}
\corr{\bm{\sigma}_{w_1}\bbr{\mbf{x}_1}\cdots \bm{\sigma}_{w_n}\bbr{\mbf{x}_n}}_{\bm{\Gamma}}  = \bigg|\prod_i w_i^{-\frac{c}{24}(w_i+1)}\,a_i^{-\frac{c}{24}\frac{w_i-1}{w_i}}\,\prod_{j} C_j^{-\frac{c}{12}} \bigg|^2\ ,
\end{equation}
\end{verticallines}
\noindent where the subscript $\bm{\Gamma}$ on the correlator denotes that this is the contribution of a single covering map, and one needs to sum over all covering maps to obtain the full correlator.
This is formally the same as the bosonic result \cite{Lunin:2000yv,Dei:2019iym}, but now the $a_i$ and $C_j$ contain also the information on the odd moduli of the insertion points $\mbf{x}_i=(x_i,\theta_i)$. By expanding in the $\theta_i$, one can extract the correlators of the superpartners of the twist fields.\footnote{For an even twist $w_i$, an extra factor $a_i^{-\frac{h_R}{w_i}}$ must be added, where $h_R$ is the dimension of the Ramond sector ground state on the covering surface.}

For two- and three-point functions, the statement in eq.~(\ref{eq:super_corr}) follows from super M\"obius Ward identities. However, for four-point functions, it is non-trivial and we have carried out explicit checks that the formula is correct for such correlators.

\begin{verticallines}[black!30!white]
%\hspace{-\parindent}
\textbf{Example.} We consider the correlator
\begin{equation}
\corr{\bm{\sigma}_{3}\bbr{\mbf{x}_1}\,\bm{\sigma}_{3}\bbr{\mbf{x}_2}\,\bm{\sigma}_{3}\bbr{\mbf{x}_3}\,\bm{\sigma}_{3}\bbr{\mbf{x}_4}}
\end{equation}
with insertion points
\begin{equation}
\mbf{x}_1 = \bbr{\infty,0}\ ,\quad \mbf{x}_2 = \bbr{2,0}\ ,\quad \mbf{x}_3 = \bbr{1,\theta_3}\ ,\quad \mbf{x}_4 = \bbr{0,\theta_4}\ .
\end{equation}
A particular super covering map for this configuration is discussed in the example around eq.~(\ref{eq:cov_3333}). Using the expansion coefficients given there, one can calculate the correlator using the super Liouville formula in eq.~(\ref{eq:super_corr}). Alternatively, one can calculate the correlator in the bosonic framework of \cite{Lunin:2000yv} by expanding the super vertex operators
\begin{equation}
\bm{\sigma}_w(\mbf{x}) = \sigma_w(x) + \theta\,G_{-\frac{1}{2}}\,\sigma_w(x)\ .
\end{equation}
The $\theta_3\theta_4$ component of the chiral part of the correlator is then given by
\begin{equation}
\theta_3\theta_4\,\corr{\sigma_3(\infty)\,\sigma_3(2)\,G_{-\frac{1}{2}}\sigma_3(1)\,G_{-\frac{1}{2}}\sigma_3(0)}\ ,
\end{equation}
which can be calculated by lifting the $G_{-\frac{1}{2}}$ mode as
\begin{align}
\cint{C(x_i)}dx\,G(x)\,\sigma_3(x_i) &\leftrightarrow \cint{C(z_i)}\frac{dz}{\sqrt{\partial\Gamma^\text{red}}(z)}\,\,G(z)\,\ket{0}\\
&= \frac{1}{\sqrt{3 a_i^\text{red}}} G(z_i)\ .\nonumber
\end{align}
On the covering surface, one is thus left with
\begin{equation}\label{eq:bos_corr_3333}
\text{prefactor}\times\theta_3\theta_4\,\frac{1}{3\sqrt{a_3^\text{red}a_4^\text{red}}}\,\corr{G(1)\,G(0)}\ ,
\end{equation}
where the prefactor is the correlator of the twist fields, and the two-point function of the supercharges is
\begin{equation}
\corr{G(1)\,G(0)} = \frac{2}{3}\,c\ .
\end{equation}
In the supersymmetric calculation, the $\theta_3\theta_4$ term instead comes from the nilpotent part of all the $a_i$ and $C_j$, given in eq.~(\ref{eq:cov_3333}). The agreement of the correlator in eq.~(\ref{eq:bos_corr_3333}) with eq.~(\ref{eq:super_corr}) is thus a non-trivial consistency check on our super covering map proposal.
\end{verticallines}

\subsubsection{Three-point functions of primaries}\label{ssec:orbifold_examples}

Let us discuss an extended example of certain three-point functions in the orbifold to see how the super covering map encodes supersymmetric information. Consider three super primary fields
\begin{equation}
\Phi_i(\mbf{x}) = \varphi_i(x) + \theta\, \psi_i (x)
\end{equation}
of dimension $h_i$, and their correlator
\begin{equation}
\mcl{C} = \corr{\mbf{V}\bbr{\varphi_{1,-\frac{h_1}{w_1}}\sigma_{w_1};\mbf{x}_1}\, \mbf{V}\bbr{\varphi_{2,-\frac{h_2}{w_2}}\sigma_{w_2};\mbf{x}_2}\, \mbf{V}\bbr{\varphi_{3,-\frac{h_3}{w_3}}\sigma_{w_3};\mbf{x}_3}}\ .
\end{equation}
The mode numbers are chosen such that the $\varphi_i$ component of the super field lifts to a primary on the covering surface. We again take all the $w_i$ odd. The case where two $w_i$ are even and we need to consider Ramond punctures is discussed in Appendix~\ref{sapp:R_3pt}.

Using super M\"obius transformations, we can fix the insertion points
\begin{equation}
\mbf{x}_1 = \bbr{\infty,0}\ ,\quad \mbf{x}_2 = \bbr{1,0}\ ,\quad \mbf{x}_3 = \bbr{0,\theta_3}\ .
\end{equation}
Let us carry out the calculation of the correlator using the super covering map $\bm{\Gamma}$ with ramification points
\begin{equation}
\mbf{z}_1 = \bbr{\infty,0}\ ,\quad \mbf{z}_2 = \bbr{1,0}\ ,\quad \mbf{z}_3 = \bbr{0,\Theta_3}\ ,
\end{equation}
and leading Taylor coefficients $a_i$ around these points. Note that, as there is only one odd modulus $\theta_3$, the $a_i$ are the same as the coefficients of $\Gamma^\text{red}$.
By the discussion around eq.~(\ref{eq:super_mode_lifting}), the fields in the correlator above are lifted as
\begin{align}
\mbf{V}\bbr{\varphi_{i,-\frac{h_i}{w_i}}\sigma_{w_i};\mbf{x}_i} &= \cint{\mbf{C}(\mbf{x}_i)}d\mbf{x}_0\,\theta_{0i}\,\mbf{x}_{0i}^{-\frac{h_i}{w_i}+h-1}\,\Phi(\mbf{x}_0)\,\bm{\sigma}_{w_i}(\mbf{x}_i)\nonumber\\
&\leftrightarrow \cint{\mbf{C}(\mbf{z}_i)}d\mbf{z}_0 \br{w_i^{1-h_i}\,a_i^{-\frac{h_i}{w_i}}\,\mbf{z}_{0i}^{-1} + \dots}\,\Phi(\mbf{z}_0)\,\ket{0}\nonumber\\
&= w_i^{1-h_i}\,a^{-\frac{h_i}{w_i}}_i\,\Phi(\mbf{z}_i)\ .
\end{align}
The correlator calculated using the super covering map is then given by
\begin{align}\label{eq:super_3_point}
\mcl{C} &= \text{prefactor} \times a_1^{-\frac{h_1}{w_1}} a_2^{-\frac{h_2}{w_2}} a_3^{-\frac{h_3}{w_3}} \times  \corr{\Phi_1(\mbf{z}_1)\,\Phi_2(\mbf{z}_2)\,\Phi_3(\mbf{z}_3)}\nonumber\\
&= \text{prefactor} \times a_1^{-\frac{h_1}{w_1}} a_2^{-\frac{h_2}{w_2}} a_3^{-\frac{h_3}{w_3}} \times  \br{C_{\varphi_1 \varphi_2 \varphi_3} + \Theta_3\,C_{\varphi_1\varphi_2\psi_3}}\ ,
\end{align}
where the prefactor contains the bare twist correlator and the $w$ dependence. To obtain the second line, we have plugged in the coordinates $\mbf{z}_i$ and introduced the $C_{abc}$ terms to denote the structure constants of the seed theory.

Let us now compare this to the bosonic calculation with the covering map $\Gamma^\text{red}$. As the superfields are not lifted as a whole but rather component-wise, we need to split up
\begin{equation}
\mbf{V}\bbr{\varphi_{i,-\frac{h_i}{w_i}}\sigma_{w_i};\mbf{x}_i} = V\bbr{\varphi_{i,-\frac{h_i}{w_i}}\sigma_{w_i};x_i} + \theta_i\,V\bbr{G_{-\frac{1}{2}}\varphi_{i,-\frac{h_i}{w_i}}\sigma_{w_i};x_i}
\end{equation}
and lift each term separately. The correlator can then be written as
\begin{equation}
\mcl{C} = \text{prefactor} \times a_1^{-\frac{h_1}{w_1}} a_2^{-\frac{h_2}{w_2}} a_3^{-\frac{h_3}{w_3}} \times  \corr{\varphi_1(\infty)\,\varphi_2(1)\,\bigg(1+\theta_3\,\cint{C(0)}\frac{dz}{\sqrt{\partial \Gamma^\text{red}}(z)}\,G(z)\bigg)\varphi_3(0)}\ .
\end{equation}
Note that the $\theta_3$ term does not simply give the superpartner $\psi_3$, because $\sqrt{\partial \Gamma^\text{red}}$ has a zero of order $\frac{w_3-1}{2}$ at $z=0$. If we expand
\begin{equation}
\frac{1}{\sqrt{\partial\Gamma^\text{red}}(z)} = \frac{b_{\frac{w_3-1}{2}}}{z^{\frac{w_3-1}{2}}} + \dots + \frac{b_1}{z} + b_0 + \mcl{O}(z)\ ,\quad z\to 0\ ,
\end{equation}
we get
\begin{equation}
\cint{C(0)}\frac{dz}{\sqrt{\partial \Gamma^\text{red}}(z)}\,G(z)\,\varphi_3(0) = \sum_{k=0}^{\frac{w_3-1}{2}}b_k\,G_{-k-\frac{1}{2}}\,\varphi_3(0)\ .
\end{equation}
The correlator can be calculated by wrapping the contour away from $z=0$. It only picks up a term at $z=1$, giving
\begin{align}
\corr{\varphi_1(\infty)\,\varphi_2(1)\,\cint{C(0)}\frac{dz}{\sqrt{\partial \Gamma^\text{red}}(z)}\,G(z)\,\varphi_3(0)} &= - C_{\varphi_1\psi_2\varphi_3} \sum_{k=0}^{\frac{w_3-1}{2}} b_k \nonumber\\
&= C_{\varphi_1\varphi_2\psi_3} \sum_{k=0}^{\frac{w_3-1}{2}} b_k\ .
\end{align}
To get the last line, we used the Ward identity $C_{\varphi_1\psi_2\varphi_3} = - C_{\varphi_1\varphi_2\psi_3}$.\footnote{This follows from applying the super translation $\tilde{z}=z+\Theta \Theta_3$, $\tilde{\Theta} = \Theta - \Theta_3$ on the covering surface.} In total, the correlator is
\begin{equation}\label{eq:bos_3_point}
\mcl{C} = \text{prefactor} \times a_1^{-\frac{h_1}{w_1}} a_2^{-\frac{h_2}{w_2}} a_3^{-\frac{h_3}{w_3}} \times  \br{C_{\varphi_1 \varphi_2 \varphi_3} + C_{\varphi_1\varphi_2\psi_3}\,\theta_3\,\sum_{k}b_k}\ .
\end{equation}
Comparing eq.~(\ref{eq:bos_3_point}) to eq.~(\ref{eq:super_3_point}), we see that
\begin{equation}
\Theta_3 = \theta_3\,\sum_{k=0}^{\frac{w_3-1}{2}} b_k\ ,
\end{equation}
and thus the location of the ramification point of the super covering map contains information about the subleading terms of $\Gamma^\text{red}$. Let us now check this on the concrete examples discussed in Section \ref{ssec:genus_zero}.

\begin{verticallines}[black!30!white]
%\hspace{-\parindent}
\textbf{Example 1.}
First, consider the ramification profile $(5,3,3)$, for which the covering map is shown in eq.~(\ref{eq:cov_533}). There,
\begin{equation}
\Theta_3 = -\sqrt{\frac{2}{15}}\,\theta_3\ ,
\end{equation}
and indeed
\begin{equation}
\frac{1}{\sqrt{\partial\Gamma^\text{red}}(z)} = -\frac{1}{\sqrt{30}\,z}-\frac{1}{\sqrt{30}} + \dots\ .
\end{equation}
\end{verticallines}

\begin{verticallines}[black!30!white]
%\hspace{-\parindent}
\textbf{Example 2.}
An interesting effect appears for ramifications $(3,3,3)$, where
\begin{equation}
\frac{1}{\sqrt{\partial\Gamma^\text{red}}(z)} = \frac{i}{\sqrt{6}\,z}-\frac{i}{\sqrt{6}} + \dots\ ,
\end{equation}
which implies $\Theta_3=0$, independently of $\theta_3$. We already observed this for the covering map in eq.~(\ref{eq:cov_333}). The special property of the $\theta_3$ part of the correlator being zero for this configuration\footnotemark~of twist operators is thus encoded in the super covering map.
\end{verticallines}
\footnotetext{This is the case for many ramification profiles. For example, it also happens for $(5,5,3)$, $(7,5,5)$, $(7,7,3)$, $(7,7,7)$, $(9,7,5)$, \dots\ . At present, we have no systematic understanding for which ramification profiles this effect occurs.}

\section{Coverings in String Theory}\label{sec:string_theory}

In this section we discuss how the structure of the super covering map can be seen on the string theory side. We first qualitatively review the results of the bosonic analysis of \cite{Eberhardt:2019ywk,Dei:2020zui} and comment on how the manifest duality with the orbifold CFT could be extended to a supersymmetric setting. We also discuss difficulties in proving such results in general, and the boons and banes of available formalisms. We then present an analysis of string theory Ward identities for genus zero worldsheets. We study these constraints in the RNS formalism \cite{Maldacena:2000hw} as well as the hybrid formalism \cite{Berkovits:1999im}, specifically in the free field realisation for $k=1$ \cite{Dei:2020zui}. In the RNS formalism, we solve the Ward identities in the case where there is no spacetime $\theta$ variable. In the hybrid formalism, we analyse correlators with non-zero spacetime $\theta$ and show that they naturally solved by the super covering map if one sets the odd worldsheet coordinate $\Theta$ to zero.

\subsection{Localisation and different formalisms}\label{ssec:expectations}

We describe how the string theory dual of the symmetric product orbifold could manifestly reproduce the CFT correlators by localising on the super moduli space, and why showing such a property of simultaneous worldsheet and spacetime supersymmetry in general is difficult due to features of the string theory formalism. We also introduce special cases of such a localisation, which we will show in the following sections.

Let us first recall how the holographic duality manifests in the case without geometric supersymmetry. The correlators of type IIB string theory on $\text{AdS}_3\times S^3\times \mbb{T}^4$ at level $k=1$ are localised on the moduli space of (bosonic) string worldsheets \cite{Eberhardt:2019ywk,Dei:2020zui}. The theory was analysed in two different formalisms; the RNS formalism \cite{Eberhardt:2019ywk,Dei:2022pkr} and the hybrid formalism \cite{Dei:2020zui}. Both formalisms have some up- and downsides. The RNS formalism \cite{Maldacena:2000hw} is a standard formulation of super string theory, which however encounters some unitarity issues at the level $k=1$ \cite{Eberhardt:2018ouy}. The analysis of the Ward identities undertaken in \cite{Eberhardt:2019ywk} has localised solutions, but also further solutions not of this form \cite{Dei:2022pkr}, and it can be argued that only including the localised solutions at $k=1$ gives a consistent theory \cite{Eberhardt:2025sbi}. The hybrid formalism, on the other hand, is well-defined at $k=1$ \cite{Eberhardt:2018ouy} and all the solutions to the Ward identities are localised \cite{Dei:2020zui}. However, the field definitions in this formalism are complicated and include a twisting of the worldsheet supersymmetry algebra \cite{Berkovits:1999im}.\\
This localisation explicitly reproduces the way correlators are calculated in the dual $\text{Sym}^N\bbr{\mbb{T}^4}$ CFT; the string worldsheet can be identified with the covering surface appearing in the orbifold \cite{Lunin:2000yv,Pakman:2009zz}. In fact, this identification can be made independently for left- and right-moving degrees of freedom. Let us briefly expand upon this, as it will be relevant for the supersymmetric generalisation. As is discussed in \cite[Section~5]{Witten:2012bg}, the bosonic worldsheet has holomorphic and antiholomorphic coordinates $(z,\tilde{z})$, but we do not necessarily need to impose that $\tilde{z}$ is the complex conjugate of $z$, $\tilde{z}\neq \bar{z}$. The string worldsheet is then a submanifold of the direct product of two Riemann surfaces $\Sigma_L^\text{red}\times\Sigma_R^\text{red}$ with coordinates $z$, $\tilde{z}$, and opposite complex structure. The analyses of \cite{Eberhardt:2019ywk,Dei:2020zui} focused only on the $z$ dependence and showed that we can identify $\Sigma_L^\text{red}$ with a covering surface appearing in the calculation of a chiral orbifold correlator, where the spacetime coordinate $x$ is given by a covering map $x=\Gamma^\text{red}(z)$. This then extends to the full correlator; if we do not fix $\tilde{z}=\bar{z}$, we are essentially considering a complexification of the spacetime, where also the boundary coordinates $x,\tilde{x}$ and their coverings are independent. Conversely, for a real spacetime, $\tilde{x}=\bar{x}$ means that we should also impose $\tilde{z}=\bar{z}$. The Ward identities then fix that the string worldsheet is the diagonal in $\Sigma_L^\text{red}\times\Sigma_R^\text{red}$, where $\Sigma_L^\text{red}=\bar{\Sigma}_R^\text{red}$ is the covering surface.\\
What can we expect if we treat supersymmetry of the CFT manifestly? As we saw in Section \ref{sec:orbifold}, correlators of twisted fields in the $\text{Sym}^N\bbr{\mbb{T}^4}$ SCFT can be written as sums over correlators on super covering surfaces. A natural analogue of the bosonic localisation could appear in the RNS formalism. As explained in \cite[Section~5]{Witten:2012bg}, the type II superstring worldsheet can be seen as a submanifold of the direct product of two SRSs $\Sigma_L\times\Sigma_R$. The local coordinates on the worldsheet are of the form $\bbr{z,\Theta;\tilde{z},\tilde{\Theta}}$, where as before we could choose $\tilde{z}=\bar{z}$, but we do not have to. On the other hand, $\Theta$ and $\tilde{\Theta}$ are not related. Analysing the Ward identities of the left-moving worldsheet correlators along the lines of \cite{Eberhardt:2019ywk,Dei:2020zui} might similarly lead to an identification of $\Sigma_L$ with a corresponding (chiral) covering SRS in the orbifold, and similarly for the right-movers. This is then consistent with a complexified super spacetime with coordinates $(x,\theta;\tilde{x},\tilde{\theta})$. To calculate the orbifold correlator on a real spacetime with $\tilde{x}=\bar{x}$, we should again impose $\tilde{z}=\bar{z}$, but keep $\theta,\tilde{\theta}$ independent.

Let us now describe how such a geometric duality could appear in RNS super string theory. Essentially, the integral over the super moduli space of worldsheets should localise to specific points describing covering surfaces. Concretely, for an $n$-point function $\langle \mbf{V}_1\cdots \mbf{V}_n\rangle$ on a genus zero worldsheet, the string theory correlator of the left-movers takes the form
\begin{equation}
\int dz_4\cdots dz_n\,\int d\Theta_3\cdots d\Theta_n \,\,\llangle \mbf{V}_1\bbr{\infty,0}\,\mbf{V}_2\bbr{1,0}\,\mbf{V}_3\bbr{0,\Theta_3}\,\prod_{j= 4}^n \mbf{V}_j\bbr{z_j,\Theta_j}\rrangle\ .
\end{equation}
Let us take $\mbf{V}_j$ as a string dual of a field in the orbifold inserted at $\mbf{x}_j=\bbr{x_j,\theta_j}$. A localised string theory correlator would then look like \cite{Eberhardt:2019ywk}
\begin{equation}\label{eq:localised_corr}
\llangle \prod_{j=1}^n \mbf{V}_j\bbr{\mbf{z}_j}\rrangle = \sum_{\bm{\Gamma}}W_{\bm{\Gamma}}\bbr{\mbf{z}_j}\,\prod_{i=4}^n \delta\bbr{z_i-z_i^{\bm{\Gamma}}}\,\prod_{j=3}^n \bbr{\Theta_j-\Theta_j^{\bm{\Gamma}}}\ ,
\end{equation}
where the sum goes over all super covering maps $\bm{\Gamma}=\bbr{\Gamma_1,\Gamma_2}$ for the orbifold correlator whose ramification points are denoted by $\mbf{z}_i^{\bm{\Gamma}}=\bbr{z_i^{\bm{\Gamma}},\Theta_i^{\bm{\Gamma}}}$. Furthermore, the factors $\bbr{\Theta_j-\Theta_j^{\bm{\Gamma}}}$ are Grassmann delta functions, and $W_{\bm{\Gamma}}$ is some unknown function.

We will not be able to show that the string theory correlators are of the localised form in eq.~(\ref{eq:localised_corr}) using the RNS formalism. We will only show a weaker property in Section \ref{ssec:RNS}, namely that eq.~(\ref{eq:localised_corr}) solves the Ward identities in the case where all $\theta_j=0$. Let us comment on the difficulties of showing a stronger localisation property. Moving the spacetime insertion point $\theta_j$ should correspond to inserting the duals of the spacetime supercharges in the string worldsheet correlator. These operators are contour integrals of spin operators on the worldsheet \cite{Giveon:1998ns}, leading to various subtleties and complicating the analytic structure of the correlators. Furthermore, the operators also affect the compact directions of the vertex operators, while one can focus only on the $\mfr{sl}\bbr{2,\mbb{R}}$ part in the bosonic case \cite{Eberhardt:2019ywk}. This makes the Ward identity analysis more involved in this case.

Some of these problems are ameliorated if one works in the hybrid formalism instead \cite{Berkovits:1999im}. In particular, setting the level to $k=1$ is well-defined and even simplifies the theory, as it admits a free field realisation \cite{Eberhardt:2018ouy,Dei:2020zui}. Furthermore, the spacetime supersymmetry corresponds to a charge in the worldsheet $\mfr{psu}(1,1|2)_1$ algebra, which makes translations in the spacetime $\theta_j$ coordinates manageable. On the other hand, there is an obstruction to realising the localisation in the form of eq.~(\ref{eq:localised_corr}), because the worldsheet supersymmetry algebra has been twisted. As a consequence, (the left-moving part of) the worldsheet is not a SRS but rather a ``semi-rigid super Riemann surface'' \cite{Distler:1991mf}. Unlike in the RNS formalism, this string worldsheet can thus not be directly identified with a covering surface in the orbifold theory. As we will show in Section \ref{ssec:hybrid}, we can still see a trace of the super covering map with a pleasing geometric interpretation by forgetting about the worldsheet supersymmetry and only including spacetime supersymmetry. Concretely, it was shown in \cite{Dei:2020zui} that the symplectic bosons of the worldsheet theory give the components of a covering map when inserted into correlators. If we make spacetime supersymmetry manifest, we see that there is a worldsheet fermion which produces a part of the odd component of the super covering map when inserted into correlators.

To summarise, the RNS and hybrid formalism are complementary in their description of worldsheet and spacetime supersymmetry. When one is manifest, the other is obscured by complicated field transformations. Considering only the easier supersymmetry geometrically, we can see special cases of the localisation in eq.~(\ref{eq:localised_corr}). This provides evidence that the super covering maps also naturally appear in string theory.

\subsection{RNS formalism}\label{ssec:RNS}

We first discuss string theory on $\text{AdS}_3\times S^3\times \mbb{T}^4$ with $k$ units of NS-NS flux and no R-R flux in the RNS formalism. We give a brief exposition to how the Ward identity set up in \cite{Eberhardt:2019ywk} (for a genus zero worldsheet) can be adapted to the supersymmetric case. We show that there is a suggestive localised solution of the form in eq.~(\ref{eq:localised_corr}) for the case where the spacetime insertion points have no odd coordinates, $\theta_j=0$.

\subsubsection{Conventions}

We will follow the conventions of \cite{Ferreira:2017pgt}, which we now briefly review. The pure NS-NS theory with $k$ units of flux is given by a supersymmetric WZW model, where the $\text{AdS}_3$ factor is described by a $\mfr{sl}\bbr{2,\mbb{R}}_k^{(1)}$ current algebra with super currents
\begin{equation}
\mbf{J}^a(\mbf{z}) = \psi^a(z) + \Theta\,J^a(z)\ ,\qquad a=+,3,-\ .
\end{equation}
These fields have an OPE
\begin{equation}
\mbf{J}^a(\mbf{z}_0)\,\mbf{J}^b(\mbf{z}_1) \sim \frac{k\,\eta^{ab}}{\mbf{z}_{01}} + \Theta_{01}\,\frac{i\,{f^{ab}}_c\,\mbf{J}^c(\mbf{z}_1)}{\mbf{z}_{01}}\ ,
\end{equation}
where $\mbf{z}_{01}=z_0-z_1-\Theta_0\Theta_1$ and $\Theta_{01}=\Theta_{0}-\Theta_1$ are super differences, and the non-zero components of the Killing form and structure constants are given by
\begin{equation}
\eta^{+-}=\eta^{-+}=-2\eta^{33}=1\ ,\qquad i\,{f^{+-}}_{3} = \mp 2 i\,{f^{3\pm}}_{\pm} = -2\ .
\end{equation}

The ground states we consider are of the form
\begin{equation}
\ket{j,m}\otimes \ket{0}_\text{NS}\ ,
\end{equation}
where $j,m$ are the quantum numbers of the $\mfr{sl}(2,\mbb{R})$ representation formed by the zero modes $J^a_0$, and we restrict to the NS sector for the fermions (the analysis of the Ward identities for the Ramond sector is similar). For $k=1$, the only $\mfr{sl}(2,\mbb{R})$ representation which appears is the principal continuous representation with $j=\frac{1}{2}$ \cite{Eberhardt:2018ouy}, but this will not play a big role in the following.

To get the dual states of twisted sectors in the orbifold, we need to spectrally flow these representations \cite{Maldacena:2000hw}. Spectral flow $\sigma$ is an outer automorphism on the algebra, such that operators act on spectrally flowed states $[\ket{\phi}]^\sigma$ as
\begin{equation}
A\,[\ket{\phi}]^\sigma = [\sigma(A)\ket{\phi}]^\sigma\ .
\end{equation}
Spectral flow acts on the fields as
\begin{align}
\sigma^w\br{\psi^\pm_{r}} &= \psi^\pm_{r\mp w}\ , & \sigma^w\br{\psi^3_{r}} &= \psi^3_{r}\ ,\nonumber\\
\sigma^w\br{J^\pm_{m}} &= J^\pm_{m\mp w}\ , & \sigma^w\br{J^3_{m}} &= J^3_{m}+\frac{k}{2}\,w\,\delta_{m,0}\ .
\end{align}
This implies that the OPE of the super current $\mbf{J}^-$ with the super vertex operators is highly regular (see Section~\ref{ssec:SCFT}),
\begin{equation}\label{eq:Jm_reg}
\mbf{J}^-(\mbf{z}_0)\,\mbf{V}\Big(\big[\ket{j,m}\otimes\ket{0}_\text{NS}\big]^{\sigma^w},\mbf{z}_1\Big) = \mcl{O}\bbr{\Theta_{01}\,\mbf{z}_{01}^{w-1}}\ .
\end{equation}
This regularity will be the basis of the power of the Ward identities, exactly as in the bosonic case \cite{Eberhardt:2019ywk}.

To get vertex operators dual to states inserted at a spacetime point $\mbf{x}=(x,\theta)$, we would have to translate the operators using a spacetime $\mfr{osp}(1|2)$ algebra \cite{Eberhardt:2019ywk}. Choosing some $\mcl{N}=1$ subalgebra of the spacetime $\mcl{N}=4$, there is identification \cite{Giveon:1998ns} with operators on the worldsheet
\begin{equation}\label{eq:rns_osp}
L^\text{ST}_{-1} = J_0^+\ ,\qquad L^\text{ST}_{0} = J_0^3\ ,\qquad L^\text{ST}_{1} = J_0^-\ ,\qquad G^\text{ST}_{-\frac{1}{2}} = S_0\ ,\qquad G^\text{ST}_{\frac{1}{2}} = S_0^\dagger\ .
\end{equation}
The spacetime supercharge $S_0$ is a very subtle operator; it is obtained by contour integrating a vertex operator composed of a field in the Wakimoto representation, and a worldsheet spin field \cite[eq.~(3.12)]{Giveon:1998ns}. Translating by $\theta$, i.e.~conjugating by $S_0$, is therefore a difficult operation on the worldsheet, which is for example not manifestly local. Furthermore, $S_0$ also affects the compact directions of $\text{AdS}_3\times S^3\times \mbb{T}^4$. In the present context, we will avoid these complications by only considering spacetime insertions at $\mbf{x}=(x,0)$ and analysing the Ward identities in this simpler case.
Describing the dual of an orbifold vertex operator at a point $\mbf{x}=(x,0)$ is achieved by translating the worldsheet vertex operator using $L^\text{ST}_{-1}$ \cite{Eberhardt:2019ywk},
\begin{equation}
\mbf{V}\bbr{\ket{\phi},x;\mbf{z}} := e^{xJ^+_0}\,\mbf{V}\bbr{\ket{\phi},\mbf{z}}\,e^{-xJ^+_0}\ .
\end{equation}

\subsubsection{Ward identities}

We look at correlators of the form
\begin{equation}
\llangle \mbf{V}_1\bbr{x_1;\mbf{z}_1} \cdots \mbf{V}_n\bbr{x_n;\mbf{z}_n} \rrangle\ ,
\end{equation}
where $\mbf{V}_i = \mbf{V}\bbr{[\ket{j_i,m_i}\otimes\ket{0}_\text{NS}]^{\sigma^{w_i}}\otimes\ket{\phi_i}_\text{cpct}}$, with $\ket{\phi_i}_\text{cpct}$ a state in the compact directions. These worldsheet correlators should reproduce those of the dual supersymmetric orbifold with operators inserted at $\mbf{x}_j=(x_j,0)$, and in particular the super covering map should play a role. However, evaluating them directly is very difficult \cite{Dei:2022pkr}.
One way to constrain these correlators is to consider current insertions and proceeding analogously to \cite{Eberhardt:2019ywk}. Let us now outline how this can be done.

The correlators are highly constrained because of the regularity, see the discussion around eq.~(\ref{eq:Jm_reg}),
\begin{equation}
\mbf{J}^-(\mbf{z}_0)\,\mbf{V}\Big(\big[\ket{j,m}\otimes\ket{0}_\text{NS}\big]^{\sigma^w},\mbf{z}_1\Big) = \mcl{O}\bbr{\Theta_{01}\,\mbf{z}_{01}^{w-1}}\ .
\end{equation}
If we consider $x$ translated operators, we need to consider ``shifted currents'', because we have OPEs
\begin{equation}
\mbf{J}^a(\mbf{z}_0)\,\mbf{V}\bbr{\ket{\phi},x;\mbf{z}_1} = e^{xJ^+_0}\,\mbf{J}^{a\,(x)}(\mbf{z}_0)\,\mbf{V}\bbr{\ket{\phi},\mbf{z}_1}\,e^{-xJ^+_0}\ ,
\end{equation}
where
\begin{equation}
\mbf{J}^{a\,(x)} = e^{-xJ^+_0}\,\mbf{J}^{a}\,e^{xJ^+_0}\ .
\end{equation}
The shifted currents are given by
\begin{align}
\mbf{J}^{+\,(x)} &= \mbf{J}^+\ ,\nonumber\\
\mbf{J}^{3\,(x)} &= \mbf{J}^3+x\,\mbf{J}^+\ ,\\
\mbf{J}^{-\,(x)} &= \mbf{J}^-+2x\,\mbf{J}^3+x^2\,\mbf{J}^+\ .\nonumber
\end{align}
Notice that we have
\begin{equation}
\br{\mbf{J}^- - 2x\,\mbf{J}^3+x^2\,\mbf{J}^+}^{(x)} = \mbf{J}^-\ .
\end{equation}
Inserting this object with different $x_j$ leads to the Ward identities \cite[Section~4]{Eberhardt:2019ywk}
\begin{equation}\label{eq:rns_ward_id}
\llangle \br{\mbf{J}^- - 2x_j\,\mbf{J}^3+x_j^2\,\mbf{J}^+}\!(\mbf{z}_0)\,\mbf{V}_1\bbr{x_1;\mbf{z}_1} \cdots \mbf{V}_n\bbr{x_n;\mbf{z}_n} \rrangle = \mcl{O}\bbr{\Theta_{0j}\,\mbf{z}_{0j}^{w_j-1}} \ ,
\end{equation}
for $\mbf{z}_0$ close to $\mbf{z}_j$.

\subsubsection{A simple localised solution}

Here we show that in the simple case where $\mbf{x}_j=\bbr{x_j,0}$, the Ward identities in eq.~(\ref{eq:rns_ward_id}) have a localised solution of the form discussed in eq.~(\ref{eq:localised_corr}). The super covering maps $\bm{\Gamma}:\mbb{CP}^{1|1}\to\mbb{CP}^{1|1}$ for this spacetime configuration are essentially given by the underlying ordinary covering map $\Gamma^\text{red}:\mbb{CP}^{1}\to\mbb{CP}^{1}$, namely
\begin{equation}
\Gamma_1\bbr{\mbf{z}} = \Gamma^\text{red}(z)\ ,\qquad \Gamma_2\bbr{\mbf{z}} = \Theta\,\sqrt{\partial \Gamma^\text{red}}(z)\ .
\end{equation}
The ramification points of $\bm{\Gamma}$ are thus $\mbf{z}_i^{\bm{\Gamma}} = \bbr{z_i^{\Gamma^{\text{red}}},0}$. The localised solution in eq.~(\ref{eq:localised_corr}) for the correlator then takes the form
\begin{align}\label{eq:simple_localised_corr}
\llangle \prod_{j=1}^n \mbf{V}_j\bbr{x_j,0;\mbf{z}_j}\rrangle &= \sum_{\bm{\Gamma}}W_{\bm{\Gamma}}\bbr{\mbf{z}_j}\,\prod_{i=4}^n \delta\bbr{z_i-z_i^{\bm{\Gamma}}}\,\prod_{j=3}^n \bbr{\Theta_j-\Theta_j^{\bm{\Gamma}}}\nonumber\\
&= \sum_{\Gamma^{\text{red}}}W_{\Gamma^{\text{red}}}\bbr{z_j}\,\prod_{i=4}^n \delta\bbr{z_i-z_i^{\Gamma^{\text{red}}}}\,\prod_{j=3}^n \Theta_j \ .
\end{align}
The last line is simply $\Theta_3\cdots\Theta_n$ times the bosonic solution of \cite{Eberhardt:2019ywk}. Indeed, one can see that by making the ansatz above, i.e.~that $\Theta_j\langle\cdots\rangle = 0$, the Ward identities reduce to those in \cite{Eberhardt:2019ywk}, and we can use the result presented there. Let us very briefly show how this works. The strategy is entirely analogous to \cite[Section~4]{Eberhardt:2019ywk}, so we will not give many details. Define the following (unknown) quantities
\begin{align}
\Psi^i_s &= \big\langle \mbf{V}(\psi^+_s\,[\ket{j_i,m_i}]^{\sigma^{w_i}},x_i;\mbf{z}_i)\,\prod_{j\neq i}\,\mbf{V}_j(x_j;\mbf{z}_j) \big\rangle\ ,\nonumber\\
F^i_m &= \big\langle \mbf{V}(J^+_m\,[\ket{j_i,m_i}]^{\sigma^{w_i}},x_i;\mbf{z}_i)\,\prod_{j\neq i}\,\mbf{V}_j(x_j;\mbf{z}_j)\big\rangle\ ,\\
\tilde{F}^i_0 &= \big\langle \mbf{V}(J^3_0\,[\ket{j_i,m_i}]^{\sigma^{w_i}},x_i;\mbf{z}_i)\,\prod_{j\neq i}\,\mbf{V}_j(x_j;\mbf{z}_j) \big\rangle\ .\nonumber
\end{align}
Setting all $\theta_j=0$, the Ward identities in eq.~(\ref{eq:rns_ward_id}) reduce to
\begin{align}\label{eq:first_ward_id}
\mcl{O}\bbr{\Theta_{0j}\,\mbf{z}_{0j}^{w_j-1}}&= \big\langle \big(\mbf{J}^- -2x_j\,\mbf{J}^3+x^2_j\,\mbf{J}^+\big)\!(\mbf{z}_0)\prod_{j\neq i}\,\mbf{V}_j(x_j;\mbf{z}_j)\big\rangle\nonumber\\
 &=\sum_{i\neq j}\sum_{s=1/2}^{w_i+1/2}\frac{-2(x_j-x_i)\,\Theta_{0i}\,\tilde{F}^i_0\,\delta_{s,1/2} + (x_j-x_i)^2\,(\Psi^i_s+\Theta_{0i}\,F^i_{s-1/2})}{\mbf{z}_{0i}^{s+1/2}} \ .
\end{align}
We now make the ansatz
\begin{equation}\label{eq:bosonic_ansatz}
F^i_m = \Theta_3\cdots\Theta_n\,F^{i,\text{bos}}_m\ ,\quad \tilde{F}^i_0 = \Theta_3\cdots\Theta_n\,\tilde{F}^{i,\text{bos}}_0\ ,\quad \Psi^i_s = 0\ ,
\end{equation}
where $F^{i,\text{bos}}_m$ are the bosonic correlators from \cite{Eberhardt:2019ywk}. This imposes the relation $\Theta_j\,F^i_m = 0$ for all $j$, and hence the equations in eq.~(\ref{eq:first_ward_id}) above further reduce to
\begin{equation}
\mcl{O}\bbr{\Theta_{0}\,(z_0-z_j)^{w_j-1}} =\sum_{i\neq j}\sum_{s=1/2}^{w_i+1/2}\frac{-2(x_j-x_i)\,\Theta_{0}\,\tilde{F}^i_0\,\delta_{s,1/2} + (x_j-x_i)^2\,\Theta_{0}\,F^i_{s-1/2}}{(z_0-z_i)^{s+1/2}} \ .
\end{equation}
Plugging in eq.~(\ref{eq:bosonic_ansatz}), we finally get
\begin{equation}
\mcl{O}\bbr{(z_0-z_j)^{w_j-1}} =\sum_{i\neq j}\sum_{s=1/2}^{w_i+1/2}\frac{-2(x_j-x_i)\,\tilde{F}^{i,\text{bos}}_0\,\delta_{s,1/2} + (x_j-x_i)^2\,F^{i,\text{bos}}_{s-1/2}}{(z_0-z_i)^{s+1/2}} \ ,
\end{equation}
which is the bosonic recursion relation \cite[eq.~(4.9)]{Eberhardt:2019ywk}. Therefore, if we know the bosonic solution, we can multiply it by $\Theta_3\cdots\Theta_n$ to obtain a solution in the supersymmetric case with all $\theta_j=0$. This tells us that the localised expression in eq.~(\ref{eq:simple_localised_corr}) indeed solves the Ward identities.

\subsection{Hybrid formalism}\label{ssec:hybrid}

We now turn to the analysis of the correlators in the hybrid formalism. Here, we can easily translate the spacetime $\theta_j$, in which case the super covering maps appearing the orbifold of Section~\ref{sec:orbifold} become non-trivial extensions of the ordinary covering map. However, as discussed in Section~\ref{ssec:expectations}, the worldsheet supersymmetry in the formalism is twisted. We thus consider no explicit worldsheet supersymmetry. Still, by extending the analysis of \cite{Dei:2020zui} to include the worldsheet fermions, we can see part of the super covering map appearing. Intuitively, the worldsheet fermion reproduces the odd component of the super covering map when inserted into correlators.

\subsubsection{Conventions}

We briefly review the free field realisation of the hybrid formalism at $k=1$, see \cite{Dei:2020zui} for the conventions we will follow. The $\mfr{psu}(1,1|2)_1$ algebra is realised by introducing symplectic boson $\xi^\pm,\eta^\pm$, i.e.~dimension $1/2$ bosons, and free fermions $\psi^\pm,\chi^\pm$. They satisfy the commutation relations
\begin{equation}
[\xi^\alpha_r,\eta^\beta_s]=\epsilon^{\alpha\beta}\,\delta_{r+s,0}\ ,\quad \{\psi^\alpha_r,\chi^\beta_s\}=\epsilon^{\alpha\beta}\,\delta_{r+s,0}\ .
\end{equation}
The Ramond sector vacuum representations we consider are spanned by $\ket{m_1,m_2}\otimes\ket{\updownarrow}$, where
\begin{align}
\xi^+_0\ket{m_1,m_2}&= \ket{m_1,m_2+1/2}\ , &\quad\eta^+_0\ket{m_1,m_2}&=2m_1\,\ket{m_1+1/2,m_2}\ ,\nonumber\\
\xi^-_0\ket{m_1,m_2}&=-\ket{m_1-1/2,m_2}\ , &\quad\eta^-_0\ket{m_1,m_2}&=-2m_2\,\ket{m_1,m_2-1/2}\ ,
\end{align}
and
\begin{equation}
\chi^+_0\ket{\uparrow}=\psi^+_0\ket{\uparrow} = 0\ ,\quad \chi^-_0\ket{\downarrow}=\psi^-_0\ket{\downarrow} = 0\ .
\end{equation}
The representations of the worldsheet theory are the spectrally flowed Ramond sector vacuum representations.
On the fermions and symplectic bosons, the spectral flow is given by
\begin{align}
\sigma^w(\xi^+_r) &= \xi^+_{r-w/2}\ ,&\quad \sigma^w(\eta^+_r)&=\eta^+_{r-w/2}\ ,\nonumber\\
\sigma^w(\xi^-_r) &= \xi^-_{r+w/2}\ ,&\quad \sigma^w(\eta^-_r)&=\eta^-_{r+w/2}\ ,\nonumber\\
\sigma^w(\psi^+_r) &= \psi^+_{r+w/2}\ ,&\quad \sigma^w(\chi^+_r)&=\chi^+_{r+w/2}\ ,\nonumber\\
\sigma^w(\psi^-_r) &= \psi^-_{r-w/2}\ ,&\quad \sigma^w(\chi^-_r)&=\chi^-_{r-w/2}\ .
\end{align}
We consider vertex operators for odd $w$
\begin{equation}
V([\ket{m_1,m_2}]^{\sigma^w}\otimes\ket{0}_\text{NS},z)\ ,
\end{equation}
which means that these states are the NS vacuum with respect to the free fermions, see \cite[Appendix A]{Dei:2020zui}.

To describe spacetime supersymmetry, we need the string theory duals of the spacetime super M\"obius generators. One can check that the following choice of worldsheet zero modes forms an $\mfr{osp}(1|2)$ algebra\footnote{There are many choices for the supercharges constructed from the $\mfr{psu}(1,1|2)$ currents, as the spacetime has $\mcl{N}=4$ supersymmetry. The most general one is $G^\text{ST}_{\mp 1/2}=\pm\bbr{a_1 S_0^{\pm ++}+a_2 S_0^{\pm -+} + b_1 S_0^{\pm +-} + b_2 S_0^{\pm --}}$ with $a_1b_2 - a_2b_1=1$. Specialising to the choice above does not affect the Ward identity analysis, see footnote \ref{footnote:different_psi}.\label{footnote:different_osp}}:
\begin{gather}\label{eq:spacetime_osp}
L^\text{ST}_{-1}= J^+_0\ ,\quad L^\text{ST}_0 = J^3_0\ ,\quad L^\text{ST}_1 = J^-_0 \ ,\nonumber\\
G^\text{ST}_{-1/2}=:S_0 = S^{+++}_0+S^{+--}_0\ ,\quad G^\text{ST}_{1/2} =:\widetilde{S}_0 = -S^{-++}_0-S^{---}_0\ .
\end{gather}
Here, $S^{\alpha\beta +}=\xi^\alpha\chi^\beta$ and $S^{\alpha\beta -} = -\eta^\alpha \psi^\beta$ are the $\mfr{psu}(1,1|2)_1$ supercurrents \cite{Dei:2020zui}.\footnote{In the full string theory, one should work with slightly different spacetime supercharges. Concretely, to satisfy physical state conditions, one has to use DDF operators, given for example in \cite[Section 4.2]{Naderi:2022bus}. These differ from the ones above by terms containing various ghost fields. However, we expect that the derivation of the Ward identities in the following section is in essence unaffected by this subtlety. In particular, the relevant shifted fields defined in eq.~(\ref{eq:shifted_hybrid_fields}) are the same. Thus the same field combinations appear in the regularity conditions, giving the same analytic structure.}

\subsubsection{Ward identities}\label{ssec:ward_identities}

We now consider shifted vertex operators dual to states inserted at a spacetime point $\mbf{x}$ with a non-zero odd coordinate $\theta$, using the spacetime supercharge $S_0$,
\begin{equation}
V(\ket{\phi},\mbf{x};z) := e^{xJ^+_0+\theta S_0}\,V(\ket{\phi},z)\,e^{-xJ^+_0-\theta S_0}\ ,
\end{equation}
and want to constrain $n$-point correlators of such operators for
\begin{equation}
\ket{\phi_i} = [\ket{m_1^i,m_2^i}]^{\sigma^w}\otimes\ket{0}_\mrm{NS}\ ,
\end{equation}
with $w_i$ odd. Correlators of these fields are dual to correlators of super twist operators in the orbifold. The OPE of the fields $\xi^+,\eta^+$ with this state has a pole of order $\frac{w+1}{2}$, while that of $\xi^-,\eta^-$ has a zero of order $\frac{w-1}{2}$. As is discussed in \cite{Dei:2020zui}, in the free field realisation, extra care needs to be taken when defining a correlator. After the dust has settled, for a genus zero worldsheet it is given by
\begin{equation}\label{eq:full_hybrid_correlator}
\Big\langle \prod_{\alpha=1}^{n-2}W(u_\alpha)\,V(\ket{\phi_1},\mbf{x}_1;z_1)\,V(\ket{\phi_2},\mbf{x}_2;z_2)\,\prod_{i=3}^n V(\mcl{Q}_{-1}\ket{\phi_i},\mbf{x}_i;z_i)\, \Big\rangle \ ,
\end{equation}
We briefly explain the extra components of this correlator. The modes of
\begin{equation}
\mcl{Q} = 2\chi^+\chi^-\,\big(\xi^+\,\partial\xi^--\partial\xi^+\,\xi^-\big)
\end{equation}
appear because of the application of super charges used in the definition of correlators in the hybrid formalism \cite{Berkovits:1999im}. This is similar to picture changing in the RNS formalism. The $W$ field is the vacuum with respect to the $\mfr{psu}(1,1|2)_1$ algebra, but is necessary to balance an extra charge in the free field realisation. Its explicit form is given in \cite{Dei:2020zui}, but we will only need the property that the OPE of $\xi^\pm$ and $\psi^\pm$ with $W$ has a first order zero.

The addition of the $\mbf{x}$ variable affects the calculation of OPEs in the following way. Bringing $A(z)$ close to $V(\ket{\phi},\mbf{x};\zeta)$ gives
\begin{equation}
A(z)\,V(\ket{\phi},\mbf{x};\zeta)=e^{xJ^+_0+\theta S_0}\,A^{(x,\theta)}(z)\,V(\ket{\phi},\zeta)\,e^{-xJ^+_0-\theta S_0}\ ,
\end{equation}
with $A^{(x,\theta)}(z)=e^{-xJ^+_0 -\theta S_0}\,A(z)\,e^{xJ^+_0+\theta S_0}$. The $\xi^\pm$ and $\psi^-$ are shifted as follows\footnote{A different choice of spacetime $\mfr{osp}(1|2)$ algebra, mentioned in footnote \ref{footnote:different_osp}, corresponds to replacing $\psi\mapsto b_1\psi^+ + b_2\psi^-$ in the shifted fields. As the OPE of this fermion with the vertex operators has the same regularity properties as $\psi^-$, the Ward identity analysis is unchanged.\label{footnote:different_psi}}:
\begin{align}\label{eq:shifted_hybrid_fields}
\xi^{-\,(x,\theta)}(z) &= \xi^-(z) + \theta\,\psi^-(z)-x\,\xi^+(z)\ , &\quad \xi^{+\,(x,\theta)}(z) &= \xi^+(z)\ ,\\
\psi^{-\,(x,\theta)}(z) &= \psi^-(z) + \theta\,\xi^+(z)\ .& &
\end{align}
Therefore, we have the relations
\begin{equation}\label{eq:shift_relations}
\big(\xi^-(z)+x\,\xi^+(z)-\theta\,\psi^-(z)\big)^{(x,\theta)} = \xi^-(z)\ ,\quad \big(\psi^-(z)-\theta\,\xi^+(z)\big)^{(x,\theta)} = \psi^-(z) \ ,
\end{equation}
which we can use to calculate Ward identities along the lines of \cite{Dei:2020zui}. We define
\begin{subequations}
\begin{align}
P^\pm(z) &= \frac{\prod_i (z-z_i)^{\frac{w_i+1}{2}}}{\prod_\alpha (z-u_\alpha)}\,  \big\langle \xi^\pm(z)\,\cdots  \big\rangle \ , \\
\Lambda(z) &= \frac{\prod_i (z-z_i)^{\frac{w_i+1}{2}}}{\prod_\alpha (z-u_\alpha)}\,\big\langle \psi^-(z)\, \cdots \big\rangle \ ,
\end{align}
\end{subequations}
where the dots denote the fields in eq.~(\ref{eq:full_hybrid_correlator}). Because the symplectic bosons and free fermions have conformal dimension $\frac{1}{2}$, $P^\pm$ and $\Lambda$ are polynomials of degree $N$, as the correlators decay as $1/z$ for $z\to \infty$. However, the coefficients are now also dependent on the $\theta_i$.

From eq.~(\ref{eq:shift_relations}), it follows that near the insertions $z_k$, these polynomials satisfy the relations
\begin{align}\label{eq:hybrid_regularity_conditions}
P^-(z)+x_i\,P^+(z)+\Lambda(z)\,\theta_i &= \mcl{O}\big((z-z_i)^{w_i}\big)\ ,\nonumber\\
\Lambda(z) - P^+(z)\,\theta_i &= \mcl{O}\big((z-z_i)^{\frac{w_i-1}{2}}\big)\ ,\quad z\to z_i\ .
\end{align}
More precisely, these relations hold in this form for $i\geq 3$. For $i=1,2$, $\Lambda-\theta_i\,P^+$ actually goes as $(z-z_i)^{\frac{w_i+1}{2}}$, because no $\mcl{Q}_{-1}$ mode is applied with which $\psi^-$ has a pole.

The regularity conditions (\ref{eq:hybrid_regularity_conditions}) look qualitatively similar to eq.~(\ref{eq:poly_reg}), and we will now see that super covering maps naturally solve these equations.

\subsubsection{Solution in terms of the super covering map}\label{ssec:ward_id_sol}

As in \cite{Dei:2020zui}, we can see that $P^\pm$ and $\Lambda$ can only be non-zero if a bosonic covering map $\Gamma^\text{red}$ exists for the choice of $z_i$.\footnote{This follows from expanding $P^\pm, \Lambda$ in the $\theta_i$, and comparing order by order; if no $\Gamma^\text{red}$ exists, the $\theta_i\theta_j$ independent components of $P^\pm$ are zero by the argument in \cite{Dei:2020zui}. It then follows that the components of $\Lambda$ which are first order in $\theta_i$ are zero by the second regularity condition. This then shows that the $\theta_i\theta_j$ components of $P^\pm$ are zero, and so on.} This implies that the correlators localise,
\begin{equation}
\Big\langle \prod_{\alpha=1}^{n-2}W(u_\alpha)\,\prod_{i=1}^n V_i(\mbf{x}_i;z_i)\, \Big\rangle = \sum_{\Gamma^\text{red}} C_{\Gamma^\text{red}}\,f(\theta_i)\,\prod_{i=4}^{n} \delta\bbr{z_i - z_i^{\Gamma^\text{red}}}\ ,
\end{equation}
where $C_{\Gamma^\text{red}}$ is some unspecified function of the even variables, and $f(\theta_i)$ some function depending on the odd spacetime variables. Furthermore, $z_i^{\Gamma^\text{red}}$ denotes a ramification point of $\Gamma^\text{red}$.

The dependence on the odd variables can be further constrained using the super covering map. Let $\bm{\Gamma}$ be a super covering map extending $\Gamma^\text{red}$, with branch points $\mbf{x}_i=\bbr{x_i,\theta_i}$ and ramification points $\mbf{z}_i = \bbr{z_i,\Theta_i}$ with $\Theta_1=\Theta_2=0$ (which can always be achieved by a super M\"obius transformation on the covering surface). As discussed in Section \ref{ssec:genus_zero}, $\bm{\Gamma}$ is of the rational form
\begin{equation}
\Gamma_1(\mbf{z}) = \frac{P(\mbf{z})}{Q(\mbf{z})}\ ,\quad \Gamma_2(\mbf{z}) = \frac{\Upsilon(\mbf{z})}{Q(\mbf{z})}\ ,
\end{equation}
satisfying the regularity condition
\begin{align}\label{eq:poly_reg_2}
P\bbr{z,\Theta}-Q\bbr{z,\Theta}\,x_i-\Upsilon\bbr{z,\Theta}\,\theta_i&=\mcl{O}\Big(\bbr{z-z_i-\Theta\Theta_i}^{w_i}\Big)\ ,\\
\Upsilon\bbr{z,\Theta}-Q\bbr{z,\Theta}\,\theta_i&=\mcl{O}\Big(\bbr{\Theta-\Theta_i}\,\bbr{z-z_i-\Theta\Theta_i}^{\frac{w_i-1}{2}}\Big)\ ,\quad i=1,\dots,n\ .\nonumber
\end{align}
By setting the worldsheet Grassmann coordinate $\Theta$ to zero, these regularity conditions reduce to eq.~(\ref{eq:hybrid_regularity_conditions}). The Ward identities can thus be solved by identifying
\begin{equation}\label{eq:integrated_super_covering}
-P^-(z) = P\bbr{z,0}\ , \quad
P^+(z) = Q\bbr{z,0}\ ,\quad
\Lambda(z) = \Upsilon\bbr{z,0}\ .
\end{equation}
The symplectic bosons $\xi^\pm$ and fermion $\psi^-$ thus behave like the $\Theta$ independent components of a super covering map expressed in projective coordinates $[x_1:x_2:\theta]$ of $\mbb{CP}^{1|1}$.

Note an interesting consequence: by comparing the leading coefficients for the odd component in eq.~(\ref{eq:hybrid_regularity_conditions}) and eq.~(\ref{eq:poly_reg_2}), one can formally extract an odd worldsheet coordinate insertion $\Theta_i$ for $i=3,\dots, n$. It would be interesting to see if there exists a formulation of the theory which makes such worldsheet supersymmetry insertions manifest, and in particular makes the $\Theta$ component of $\bm{\Gamma}$ visible in string theory correlators.

\subsubsection{Explicit solutions}
As in \cite[Section 5]{Dei:2020zui}, we can solve the Ward identities explicitly, to check in specific examples that the description (\ref{eq:integrated_super_covering}) is correct. We are again interested in the correlator
\begin{equation}
\Big\langle \prod_{\alpha=1}^{n-2}W(u_\alpha)\,V(\ket{\phi_1},\mbf{x}_1;z_1)\,V(\ket{\phi_2},\mbf{x}_2;z_2)\,\prod_{i=3}^n V(\mcl{Q}_{-1}\ket{\phi_i},\mbf{x}_i;z_i)\, \Big\rangle \ .
\end{equation}
Using Liouville's theorem and the decay of the correlators $\langle \xi^\pm \cdots\rangle$ for $z\to \infty$, we know that the correlators are given by the expansion around the poles. Using then eqs.~(\ref{eq:shifted_hybrid_fields}) and the regularity of $\xi^-$ with $[\ket{m_1,m_2}]^{\sigma^w}$, one finds
\begin{subequations}
\begin{align}
\big\langle \xi^+(z)\cdots\big\rangle &= \sum_{i=1}^n\sum_{l=1}^{\frac{w_i+1}{2}}\frac{F^i_{l-1/2}}{(z-z_i)^l}\ ,\\
\big\langle \psi^-(z)\cdots\big\rangle &= \sum_{i=1}^n\sum_{l=1}^{\frac{w_i+1}{2}}\frac{\theta_i\,F^i_{l-1/2}}{(z-z_i)^l}+\sum_{i=3}^n\frac{\Psi^i}{z-z_i}\ ,\\
\big\langle \xi^-(z)\cdots\big\rangle &= -\sum_{i=1}^n\sum_{l=1}^{\frac{w_i+1}{2}}\frac{x_i\,F^i_{l-1/2}}{(z-z_i)^l}+\sum_{i=3}^n\frac{\theta_i\,\Psi^i}{z-z_i}\ .
\end{align}
\end{subequations}
Here, $F^i_s$ denotes the correlator where the mode $\xi^+_s$ is applied to the $i$'th vertex operator, and $\Psi^i$ the correlator where $\psi^-_{1/2}$ is applied. $\Psi^i$ is only non-zero for $i=3,\dots,n$ because of the presence of $\mcl{Q}_{-1}$. The regularity conditions (\ref{eq:hybrid_regularity_conditions}) translate into the conditions
\begin{equation}
\sum_{i\neq j}\sum_{l=1}^{\frac{w_i+1}{2}}\frac{(x_j-x_i-\theta_j\theta_i)\,F^i_{l-1/2}}{(z-z_i)^l} - \sum_{i\neq j,1,2}\frac{(\theta_j-\theta_i)\,\Psi^i}{z-z_i} = \mcl{O}\big((z-z_j)^{\frac{w_j-1}{2}}\big)\ ,
\end{equation}
for $j=1,\dots,n$. Furthermore, all the correlators above have first order zeros at $u_\alpha$, $\alpha=1,\dots,n-2$, because the OPE of $\xi^\pm,\psi^-$ with $W$ has a zero. Therefore, we get the additional conditions
\begin{subequations}
\begin{align}
0 &= \sum_{i=1}^n\sum_{l=1}^{\frac{w_i+1}{2}}\frac{F^i_{l-1/2}}{(u_\alpha-z_i)^l}\ ,\\
0 &= \sum_{i=1}^n\sum_{l=1}^{\frac{w_i+1}{2}}\frac{\theta_i\,F^i_{l-1/2}}{(u_\alpha-z_i)^l}+\sum_{i=3}^n\frac{\Psi^i}{u_\alpha-z_i} \ , \label{equ:psi_u_condition} \\
0 &= -\sum_{i=1}^n\sum_{l=1}^{\frac{w_i+1}{2}}\frac{x_i\,F^i_{l-1/2}}{(u_\alpha-z_i)^l}+\sum_{i=3}^n\frac{\theta_i\,\Psi^i}{u_\alpha-z_i} \ ,
\end{align}
\end{subequations}
for $\alpha = 1,\dots,n-2$. As there are only $n-2$ unknowns $\Psi^i$, eq.~(\ref{equ:psi_u_condition}) can be used to eliminate them.

Let us now discuss specific three-point functions. We fix
\begin{equation}
z_1=x_1=0\ , \quad z_2 = x_2 = 1\ ,\quad z_3=x_3 = -1 \ ,\quad \theta_1=\theta_2 = 0\ .
\end{equation}
\begin{verticallines}[black!30!white]
%\hspace{-\parindent}
\textbf{Example 1.}
Considering the ramification profile $(5,3,3)$, one obtains the unique solution
\begin{equation}\label{eq:hyb_533}
-\frac{\langle\xi^-(z)\cdots\rangle}{\langle\xi^+(z)\cdots\rangle} = \frac{8\, z^5}{15\, z^4 - 10\, z^2 + 3}\ ,\quad \frac{\langle\psi^-(z)\cdots\rangle}{\langle\xi^+(z)\cdots\rangle} = -\frac{2\, (z-1)^2 \,z^3\, \theta_3}{15\, z^4 - 10 \,z^2 + 3} \ .
\end{equation}
The super covering map $\bm{\Gamma}$ for this configuration can be found by M\"obius transforming\footnotemark~the map in eq.~(\ref{eq:cov_533}). Explicitly, it is given by
\begin{align}
\Gamma_1(\mbf{z}) &= \frac{P(\mbf{z})}{Q(\mbf{z})} = \frac{8\, z^5}{15\, z^4 - 10\, z^2 + 3+\frac{15}{2}\,(z-1)^3\,(z+1)\,\Theta\,\Theta_3}\ ,\\
\Gamma_2(\mbf{z}) &= \frac{\Upsilon(\mbf{z})}{Q(\mbf{z})}= \frac{\sqrt{30}\,z^2(z-1)\big(-2(z+1)\,\Theta +z(z-1)\,\Theta_3\big)}{15\, z^4 - 10\, z^2 + 3+\frac{15}{2}\,(z-1)^3\,(z+1)\,\Theta\,\Theta_3}\ ,\nonumber
\end{align}
with $\Theta_3 = -\sqrt{\frac{2}{15}}\,\theta_3$. Setting $\Theta=0$ as in eq.~(\ref{eq:integrated_super_covering}) then clearly reproduces eq.~(\ref{eq:hyb_533}).
\end{verticallines}
\footnotetext{In eq.~(\ref{eq:cov_533}), we chose $z_1=x_1=\infty$, $z_2=x_2=1$, $z_3=x_3=0$. The M\"obius transformation relating these configurations is $\gamma(z) = \frac{1}{2}\frac{z+1}{z}$.}
\begin{verticallines}[black!30!white]
%\hspace{-\parindent}
\textbf{Example 2.}
For ramifications $(3,3,3)$, the solution to the Ward identities is
\begin{equation}\label{eq:hyb_333}
-\frac{\langle\xi^-(z)\cdots\rangle}{\langle\xi^+(z)\cdots\rangle} = \frac{8\, z^3}{3\, z^4 + 6\,z^2 - 1}\ ,\quad \frac{\langle\psi^-(z)\cdots\rangle}{\langle\xi^+(z)\cdots\rangle} = \frac{2\, (z-1)^2\, z^2\,\theta_3}{3\, z^4 + 6\,z^2 - 1} \ .
\end{equation}
The M\"obius transformed covering map from eq.~(\ref{eq:cov_333}) is
\begin{align}
P(\mbf{z}) &= 8\, z^3 - 4\sqrt{\frac{2}{3}}i\,z^3\,\Theta\,\theta_3\ ,\nonumber\\
Q(\mbf{z}) &= 3\, z^4 + 6\,z^2 - 1 -i\sqrt{\frac{2}{3}}\,\bbr{3\,z^3 + 3\,z^2 + 3\,z -1}\,\Theta\,\theta_3\ ,\\
\Upsilon(\mbf{z}) &= 2\,z\,(z-1)\,\Big(-i\sqrt{6}\,(z+1)\,\Theta + z\,(z-1)\,\theta_3\Big)\ ,\nonumber
\end{align}
with $\Theta_3=0$. Setting $\Theta=0$ again gives the same map as in eq.~(\ref{eq:hyb_333}).
\end{verticallines}

\section{Conclusion and Outlook}\label{sec:discussion}

In this paper, we have introduced the notion of super covering maps. These are maps between super Riemann surfaces which satisfy a superconformality condition eq.~(\ref{eq:gamma_sc}) and have a simple local structure around their ramification points, see eq.~(\ref{eq:gamma_reg}). We first studied the properties of these interesting geometric objects. Then, we showed that these maps play an important role in the $\text{AdS}_3/\text{CFT}_2$ correspondence. Concretely, they appear in the symmetric product orbifold of a SCFT, where they can be used to calculate correlators in a manifestly supersymmetric way.
This manifestly supersymmetric formalism using super covering maps could be applied in the future to investigate properties of certain symmetric orbifold correlators. In particular, one could study extremal correlators and their special properties in this framework \cite{Pakman:2009ab}.\footnote{I thank Lior Benizri for discussions about this application.}

On the string theory side of the duality, we observed signs of the super covering map in the RNS and hybrid formalism. In the RNS formalism, we set up Ward identities and saw that they have a localised solution in the simple case where the odd spacetime coordinates are zero. In the hybrid formalism, super covering maps solve the Ward identities of the worldsheet correlators. The symplectic bosons and free fermions thus give components of the super covering map when inserted into correlators.

In string theory, we were unable to see the full super covering map in the correlators because of difficulties in formulating theories with simultaneous worldsheet and spacetime supersymmetry. Indeed, in the hybrid formalism analysis of Section \ref{ssec:hybrid}, only a part of the super covering map appears, because the worldsheet fields are inserted without a superpartner. As the $\mcl{N}=4$ algebra in the hybrid formalism is twisted, the natural superspace in this case is semi-rigid geometry \cite{Distler:1991mf}. This geometry does not naturally fit together with the spacetime SRS. Conversely, in the RNS formalism, it is easy to introduce worldsheet superpartners, but adding a dependence on an odd spacetime variable inserts twist fields into the correlators \cite{Giveon:1998ns}, making the Ward identity analysis more difficult. A more extensive study of these constraints would certainly be interesting.

It would be interesting to see if there exists a formulation of the tensionless worldsheet theory which could incorporate both supersymmetries manifestly, or at least make it manifest on the level of the correlators. Since we showed that super covering maps describe the supersymmetric orbifold, as well as limits of reduced supersymmetry in string theory, the appearance of super covering maps in such a formulation seems very likely. They could therefore serve as guidance for the construction of such a formulation.

\subsection*{Acknowledgments}
I thank Konstantin Baune, Lior Benizri, Bob Knighton, Shota Komatsu, Nathan McStay, Kiarash Naderi, Vit Sriprachyakul for discussions, and especially Matthias Gaberdiel for many useful conversations and suggesting this topic to me. I thank the referee for insightful comments and suggestions for improvements. I also thank Matthias Gaberdiel, Bob Knighton, Nathan McStay and Vit Sriprachyakul for comments on a draft version of this article. This work is supported through a grant from the Swiss National Science Foundation as well as the NCCR SwissMAP that is also funded by the Swiss National Science Foundation.

\appendix

\section{More on the structure of super covering maps}\label{app:structure}

In this appendix, we discuss the properties of super covering maps further. We show that there are always local coordinates in which $\bm{\Gamma}$ is simple. We also give an example of a mapping which respects the superconformal structure but is not a super covering map.

\subsection{Local expression}\label{sapp:local_expression}

Let $\bm{\Gamma}:\Sigma_1\to\Sigma_2$ be a super covering map. Here we show that around a ramification point $\mbf{z}_1$ with ramification index $w$, there exist local coordinates such that
\begin{equation}\label{eq:local_expr}
\Gamma_1(\mbf{z}_0) = \mbf{z}_{01}^w\ ,\qquad \Gamma_2(\mbf{z}_0) = \Theta_{01}\,\mbf{z}_{01}^{\frac{w-1}{2}}\ .
\end{equation}
Conversely, a map satisfying this condition is a super covering map.

To see that eq.~(\ref{eq:local_expr}) holds, we can first supertranslate the charts on $\Sigma_1$ and $\Sigma_2$ so that $\mbf{z}_1=(0,0)$, as well as $\mbf{x}_1=\bm{\Gamma}(\mbf{z}_1) = (0,0)$. The condition eq.~(\ref{eq:gamma_reg}) that $\mbf{z}_1$ is a ramification point then becomes
\begin{equation}
\Gamma_1(\mbf{z}) =: f(z) + \Theta\,\rho(z) = a\,z^w + \Theta\,\alpha\,z^{w} + b\,z^{w+1} + \dots \ ,\quad z\to 0\ ,
\end{equation}
where $a$ is not nilpotent. The root $f^{\frac{1}{w}}(z)$ exists on a neighbourhood of $z=0$ and can be inverted by the inverse function theorem. Explicitly, there is a change of coordinates $z=g(u)$, such that $f\circ g(u) = u^w$ \cite{Cavalieri_Miles_2016}. This can be extended to a superconformal change of coordinates
\begin{equation}
\bbr{z,\Theta} = \bbr{g(u),\Phi\,\sqrt{\partial g}(u)}\ .
\end{equation}
such that in the $\bbr{u,\Phi}$ coordinates
\begin{equation}
\Gamma_1\bbr{u,\Phi} = u^w + \Phi\,\sigma(u)\ ,\qquad \sigma(u) = \rho\bbr{g(u)}\,\sqrt{\partial g}(u)\ .
\end{equation}
Now define coordinates $\bbr{v,\Xi}$ by
\begin{equation}
v\bbr{u,\Phi} = u + \Phi\,\beta(u)\ ,\quad \Xi\bbr{u,\Phi} = \beta(u) + \Phi\,\Big(1+\tfrac{1}{2}\beta(u)\,\partial\beta(u)\Big)\ ,\quad \beta(u) = \frac{\sigma(u)}{w\,u^{w-1}}\ .
\end{equation}
Note that $\beta$ is well-defined because $\sigma(u) = \sigma_0\,u^w + \dots$ around $u=0$. This change of coordinates is superconformal, and
\begin{equation}
\Gamma_1\bbr{v,\Xi} = v^w\ .
\end{equation}
By eq.~(\ref{eq:gamma_sc}), the second component is then
\begin{equation}
\Gamma_2\bbr{v,\Xi} = \sqrt{w}\,\Xi\,v^{\frac{w-1}{2}}\ .
\end{equation}
This shows eq.~(\ref{eq:local_expr}).

\subsection{A non-example}\label{sapp:non_example}

Not every superconformal map $\mbf{F}:\Sigma_1\to\Sigma_2$ is a super covering map. The regularity conditions at a ramification point eq.~(\ref{eq:gamma_reg}) are important for a sensible ramification structure, as we will now discuss on a non-example.

First, let us discuss the ramification structure of a super covering map. By the arguments in the previous section, $\bm{\Gamma}$ is locally of the form
\begin{equation}
\Gamma_1\bbr{z,\Theta} = z^w\ ,\qquad \Gamma_2\bbr{z,\Theta} = \sqrt{w}\,\Theta\,z^{\frac{w-1}{2}}\ .
\end{equation}
The ramification point is at $\mbf{z}_1=(0,0)$, but this is not the only point where $D\Gamma_2$ vanishes. Indeed, the derivative at a different point $\mbf{z}_2=\bbr{z_2,\Theta_2}$ is
\begin{equation}
D\Gamma_2(\mbf{z}_2) = \sqrt{w}\,z_2^{\frac{w-1}{2}}\ ,
\end{equation}
which is zero if $z_2$ is nilpotent\footnote{The even coordinate $z_2$ of an NS puncture can also depend on the odd moduli and thus be nilpotent without being zero.} of order $\frac{w-1}{2}$ and $\Theta_2$ is arbitrary. This means that $\mbf{z}_2$ is infinitesimally close to $(0,0)$. We can still call $(0,0)$ ``the'' ramification point, however, because it is the unique point with maximal ramification index $w$, i.e.~$D^k\Gamma_2(\mbf{z}_1) = 0$ for $k=1,\dots,w-1$. Furthermore, $\bm{\Gamma}$ is only ramified over the branch point $\mbf{x}_1 = (0,0)$: if $\mbf{z}_2$ is another point where $D\Gamma_2$ vanishes, then as above $z_2^{\frac{w-1}{2}}=0$ and thus
\begin{equation}
\mbf{x}_2 = \bm{\Gamma}(\mbf{z}_2) = \Big(z_2^w,\sqrt{w}\,\Theta_2\,z_2^{\frac{w-1}{2}}\Big) = (0,0) = \mbf{x}_1.
\end{equation}
The uniqueness of the branch point and ramification point with maximal ramification index $w$ are needed for a good definition of correlators in the context of the orbifold.

Let us now show in a non-example that the sensible ramification structure of a super covering map is not satisfied by every superconformal mapping. Consider the map $\mbf{F}:~\mbb{CP}^{1|1}\to\mbb{CP}^{1|1}$ given by
\begin{equation}\label{eq:non_example}
F_1\bbr{z,\Theta} = \frac{1}{5}\,z^5 + \Theta\,\alpha\,z^3\ ,\qquad F_2\bbr{z,\Theta} = \Theta\,z^2 + \alpha\,z\ .
\end{equation}
One can check that this map satisfies $DF_1 = F_2\,DF_2$. The underlying covering map is $F^\text{red}(z) = \frac{z^5}{5}$. While $F^\text{red}$ has a ramification point of index 5 at $z=0$, $\mbf{F}$ has no such ramification point in the sense of eq.~(\ref{eq:gamma_reg}). Indeed, $\mbf{z}_1$ being a ramification point of index 5 is equivalent to $D^kF_2(\mbf{z}_1)=0$ for $k=1,\dots,4$. It is straight-forward to check that there is no $\mbf{z}_1$ satisfying these conditions if $\alpha\neq 0$. There is no unique ramification point of maximal ramification index, contrary to super covering maps. In addition, this has an effect on the branching structure of $\mbf{F}$; it has infinitely many branch points $\mbf{x}_i$, which can be seen as follows. The ramification points $\mbf{z}_1$ are characterised by
\begin{equation}\label{eq:special_ram_pt}
0 = D\Gamma_2\bbr{z_1,\Theta_1} = z_1^2 + \Theta_1\,\alpha\ ,
\end{equation}
and so a ramification point is mapped to
\begin{equation}
\mbf{x}_1 = \mbf{F}(\mbf{z}_1) = \Big(-\tfrac{4}{5}\,z_1^5, \alpha\,z_1\Big)\ .
\end{equation}
There is thus not a unique branch point; for example, one can take $\mbf{z}_1 = \bbr{\beta_1\beta_2,0}$ for odd $\beta_i$, which solves eq.~(\ref{eq:special_ram_pt}), for which $\mbf{x}_1 = \bbr{0,\alpha\beta_1\beta_2}$.

In the context of superanalytic maps, the ramification condition eq.~(\ref{eq:gamma_reg}) thus contains additional information and distinguishes a subset of maps, namely those which can be locally written as in eq.~(\ref{eq:local_expr}). This is in contrast to the underlying covering maps; there, all that is required is that $\Gamma^\text{red}$ is holomorphic, from which the local structure then follows directly \cite{Cavalieri_Miles_2016}.

\section{Covering maps with Ramond punctures}\label{app:R_cov_map}

In this appendix, we discuss the case when the covering map has ramification points with even ramification index $w\in 2\,\mbb{Z}$. We give examples and show how orbifold correlators can be calculated using these maps.

The interpretation of Ramond (R) punctures and their explicit description are discussed in \cite{Witten:2012ga}. We outlined the main complications arising when introducing such punctures in Section \ref{ssec:R_punctures}. In particular, one can work either with a multi-valued odd coordinate $\Theta^\mrm{mv}$ and a normal superderivative, or with a single-valued odd coordinate $\Theta^\mrm{sv}$ and a degenerate superderivative. We now discuss an example to show what this means in practice.

\subsection{A covering map}

Let us consider the super covering map for the correlator
\begin{equation}
\corr{\bm{\sigma}_{w+1}\bbr{\infty,0}\,\bm{\sigma}_{w}\bbr{1,0}\,\bm{\sigma}_2\bbr{0,\theta_3}}\ ,\qquad w\in 2\,\mbb{Z}\ .
\end{equation}
The underlying reduced map is
\begin{equation}\label{eq:cov_w_R_red}
\Gamma^\text{red}(z) = 1-\bbr{1+w\,z}\,\bbr{z-1}^w\ ,
\end{equation}
with ramification points $z_1=\infty$, $z_2=1$, $z_3=0$. The square root of its derivative, $\sqrt{\partial \Gamma^\text{red}}$ has a branch cut from $z=0$ to $z=1$. In order for $\bm{\Gamma}$ to be single-valued, $\Theta^\mrm{mv}$ needs to cancel the monodromy and thus also has the same branch cut. It is related to the single-valued coordinate by
\begin{equation}\label{eq:sv_mv}
\Theta^\mrm{mv} = \sqrt{z\bbr{z-1}}\,\Theta^\mrm{sv}\ .
\end{equation}
Using the multi-valued $\Theta^\mrm{mv}$, the super covering map for the configuration above is given by
\begin{align}\label{eq:cov_R_w}
\Gamma_1(\mbf{z}) &= 1-\bbr{1+w\,z}\,\bbr{z-1}^w + \sqrt{w(w+1)}\,\Theta^\mrm{mv}\,\theta_3\,\sqrt{z\,\bbr{1-z}^{2w-1}} \ , \nonumber\\
\Gamma_2(\mbf{z}) &= \sqrt{w(w+1)}\,\Theta^\mrm{mv}\,\sqrt{z\,\bbr{1-z}^{w-1}} + \theta_3\,\bbr{1-z}^{\frac{w}{2}}\ .
\end{align}
Let us discuss the ramification structure of this super covering map. Near $z=0$, one finds
\begin{equation}
\Gamma_2(\mbf{z})-\theta_3 = \sqrt{w(w+1)}\,\bbr{\Theta^\mrm{mv} - \Theta_3^\mrm{mv}}\,\sqrt{z} + \dots\ ,
\end{equation}
where
\begin{equation}\label{eq:th3_mv}
\Theta_3^\mrm{mv}=(-1)^{\frac{w}{2}}\,\sqrt{\frac{w}{4\,(w+1)}}\,\sqrt{z\bbr{z-1}}\ .
\end{equation}
Expanding around $z_2=1$, one finds similarly\footnote{The automorphism group of a SRS with two R punctures is $1|1$ dimensional, as is explained in \cite{Witten:2012ga}, and we effectively label a point on the R puncture. In contrast to the case for NS punctures, we can thus only set one $\Theta_i^\mrm{mv}$ to $0$, which we choose to be $\Theta_1^\mrm{mv}$.}
\begin{equation}\label{eq:th2_mv}
\Theta_2^\mrm{mv} = -i\sqrt{\frac{1}{w(w+1)}}\,\theta_3\,\sqrt{z\bbr{z-1}}\ .
\end{equation}
If one works with $\Theta^\mrm{mv}$, the coordinates $\Theta_i^\mrm{mv}$ of the ramification points on the covering surface are thus multi-valued functions themselves. They are associated to a single-valued modulus $\Theta_i^\mrm{sv}$ using eq.~(\ref{eq:sv_mv}).

Alternatively, one can work with the global single-valued $\Theta^\mrm{sv}$, related by eq.~(\ref{eq:sv_mv}). Changing coordinates, the super covering map is manifestly single-valued
\begin{align}
\Gamma_1\bbr{\mbf{z}^\mrm{sv}} &= 1-\bbr{1+w\,z}\,\bbr{z-1}^w + \sqrt{w(w+1)}\,\Theta^\mrm{sv}\,\theta_3\,z\,\bbr{1-z}^{w-1} \ , \nonumber\\
\Gamma_2\bbr{\mbf{z}^\mrm{sv}} &= \sqrt{w(w+1)}\,\Theta^\mrm{sv}\,z\,\bbr{1-z}^{\frac{w}{2}} + \theta_3\,\bbr{1-z}^{\frac{w}{2}}\ .
\end{align}
It is superconformal with respect to the superderivative in the presence of R punctures,
\begin{equation}
D^\mrm{sv}\Gamma_1 = \Gamma_2\,D^\mrm{sv}\Gamma_2\ ,\qquad D^\mrm{sv} = \frac{\partial}{\partial \Theta^\mrm{sv}} + z\,\bbr{z-1}\,\Theta^\mrm{sv}\,\frac{\partial}{\partial z} \ .
\end{equation}

\subsection{Mode lifting}

As we saw above, we can either write the covering map in a single-valued form or use a multi-valued $\Theta^\mrm{mv}$ coordinate. In the context of a SCFT, it is more natural to use the latter description. The reason is that if we want to write local operators for super fields in the presence of R punctures, the field should have a definite monodromy. Explicitly, if we consider the super field
\begin{equation}
\phi(\mbf{z}) = v(z) + \Theta^\mrm{mv}\,\rho(z)\ ,
\end{equation}
where $v$ is bosonic and $\rho$ fermionic, the whole object is single-valued when encircling a R puncture, because $\rho$ is multi-valued around a spin field. Similarly, a super field with fermionic bottom component picks up a minus sign when encircling a R puncture.

Let us now discuss how fields are lifted in the orbifold of a SCFT. This procedure is a bit more complicated than in the case of NS punctures, but the end result again has a simple form. We will see below that super vertex operators with even twist, inserted at $\mbf{x}_1$, are lifted to Ramond super vertex operators inserted ``at'' the Ramond puncture $\mbf{z}_1$. Concretely, if $\mbf{z}_1$ has a single-valued modulus $\Theta_1^\mrm{sv}$, inserting a Ramond super vertex operator at that point is defined as
\begin{equation}
\mbf{V}_R\bbr{\ket{\phi},\mbf{z}_1} := V\bbr{\ket{\phi},z_1} + \Theta_1^\mrm{sv}\,V\bbr{G_0\ket{\phi},z_1}\ .
\end{equation}
Note that $G_0$ is applied (to be contrasted with the application of $G_{-\frac{1}{2}}$ for NS punctures), because $\ket{\phi}$ is a state in the Ramond sector on the covering surface. At first sight, it is a bit puzzling that the single-valued odd coordinate appears in this expression, while the multi-valued one appears in super fields. The reason for this is that $\mbf{V}_R$ is a spin field and cannot be treated on the same footing as an ordinary super field, because the supercharge $G$ becomes multi-valued in its presence \cite{Friedan:1986rx}.

To see that this is a sensible and consistent prescription, we need to first define the lifting of a twist operator $\bm{\sigma}_w(\mbf{x}_1)$ for even $w$. Let $\mbf{z}_1$ be the ramified pre-image. The natural definition is that this twist corresponds to an insertion of a Ramond vacuum state $\ket{h_R}$ at $\mbf{z}_1$ on the covering surface \cite{Lunin:2001pw}. Since this state satisfies $G_0\ket{h_R}=0$, the super vertex operator has actually no $\Theta_1^\mrm{sv}$ component,
\begin{equation}
\mbf{V}_R\bbr{\ket{h_R},\mbf{z}_1} = V\bbr{\ket{h_R},z_1}\ .
\end{equation}

If we now want to lift a mode of a super field $\phi=v+\theta\,\rho$ of dimension $h$, we proceed as in Section \ref{ssec:lifting}. A fractional mode
\begin{equation}
\mbf{V}\bbr{v_{\frac{r}{w}}\sigma_w,\mbf{x}_1} = \underset{\mbf{C}(\mbf{x}_1)}{\oint} \!\! d\mbf{x}_0\,\,\theta_{01}\,\mbf{x}_{01}^{\frac{r}{w}+h-1}\,\phi(\mbf{x}_0)\,\bm{\sigma}_w(\mbf{x}_1)
\end{equation}
gets lifted to
\begin{equation}
\underset{\mbf{C}(\mbf{z}_1)}{\oint} \!\! d\mbf{z}_0\,\,\bbr{D\Gamma_2}^{1-2h}\,\bbr{\Gamma_2-\theta_1}\,\bbr{\Gamma_1-x_1-\Gamma_2\,\theta_1}^{\frac{r}{w}+h-1}\,\Phi(\mbf{z}_0)\,V\bbr{\ket{h_R},z_1}\ ,
\end{equation}
where $\Phi=\mcl{V}+\Theta^\mrm{mv}\,\mcl{P}$ is the lift of $\phi$ to the covering surface. We express the covering map in terms of $\Theta^\mrm{mv}$ and integrate over this coordinate in the Berezin integral. When using the regularity in eq.~(\ref{eq:gamma_reg}) and carrying out the Berezin integral, the leading term in the $z$ expansion that is left is proportional to
\begin{equation}\label{eq:R_puncture_lifting}
\cint{C(z_1)}dz\,\Big(\bbr{z-z_1}^{h+r-1}\,\mcl{V}(z) + \Theta^\mrm{mv}_1\,\bbr{z-z_1}^{h+r-1}\,\mcl{P}(z)\Big)\,V\bbr{\ket{h_R},z_1} \ .
\end{equation}
Since $\mcl{V}$ and $\mcl{P}$ have opposite statistics, one of them is multi-valued in the presence of the spin field $V\bbr{\ket{h_R},z_1}$, while the other is single-valued. Therefore, the integral above only makes sense for $\mcl{V}$. To obtain a sensible integrand for $\mcl{P}$, we need to change back to the single-valued modulus $\Theta_1^\mrm{sv}$, which is related to $\Theta^\mrm{mv}_1$ as in eq.~(\ref{eq:sv_mv}),
\begin{equation}
\Theta_1^\mrm{mv} = \Theta_1^\mrm{sv}\,\sqrt{z-z_1} + \dots \ ,\qquad z\to z_1\ .
\end{equation}
When plugging this in, the integral in eq.~(\ref{eq:R_puncture_lifting}) evaluates to
\begin{equation}
V\bbr{\mcl{V}_{r}\ket{h_R},z_1} + \Theta_1^\mrm{sv}\,V\bbr{\mcl{P}_{r}\ket{h_R},z_1} = \bbr{1+\Theta_1^\mrm{sv}\,G_0}\,V\bbr{\mcl{V}_{r}\ket{h_R},z_1} = \mbf{V}_R\bbr{\mcl{V}_r\ket{h_R},\mbf{z}_1}\ .
\end{equation}
When applying further fractional modes, the lifting proceeds iteratively and works in the same way.

\subsection{A three-point correlator}\label{sapp:R_3pt}

We now discuss three-point functions of primaries as an example to see that this lifting procedure gives the correct correlators. This is analogous to the example in Section \ref{ssec:orbifold_examples}. Consider three super primaries $\Phi_i = \varphi_i+\theta\,\psi_i$ and the correlator
\begin{equation}
\mcl{C} = \corr{\mbf{V}\bbr{\varphi_{1,-\frac{h_1}{w_1}}\sigma_{w_1};\infty,0}\, \mbf{V}\bbr{\varphi_{2,-\frac{h_2}{w_2}}\sigma_{w_2};1,0}\, \mbf{V}\bbr{\varphi_{3,-\frac{h_3}{w_3}}\sigma_{w_3};0,\theta_3}}\ ,
\end{equation}
where $w_2$ and $w_3$ are even. We can again compare the calculation using the super covering map and the ordinary (reduced) covering map explicitly. Using the super covering map, the correlator is given by
\begin{equation}
\mcl{C} = \text{factors}\times\corr{\mbf{V}\bbr{\varphi_{1,-h_1}\ket{0};\infty,0}\, \mbf{V}_R\bbr{\varphi_{2,-h_2}\ket{h_R};1,\Theta_2^\mrm{sv}}\, \mbf{V}_R\bbr{\varphi_{3,-h_3}\ket{h_R};0,\Theta_3^\mrm{sv}}}\ .
\end{equation}
The $\theta_3$ component of this expression is given by
\begin{equation}\label{eq:3_point_R_susy}
\Theta_2^\mrm{sv}\,\corr{V_1\,\bbr{G_{0}V_2}\,V_3} + \Theta_3^\mrm{sv}\,\corr{V_1\,V_2\,\bbr{G_{0}V_3}}
\end{equation}
with $V_j$ denoting the bottom component of the $j$'th operator above.

If one uses the reduced covering map $\Gamma^\text{red}$ instead, the $\theta_3$ component of the correlator is
\begin{equation}\label{eq:3_point_R_bos}
\theta_3\,\cint{C(0)}\frac{dz}{\sqrt{\partial \Gamma^\text{red}}(z)}\,\corr{V_1\,V_2\,G(z)\,V_3}\ .
\end{equation}
To relate the two expressions, we need to calculate the correlator with the $G$ insertion,
\begin{equation}
g(z) = \corr{V_1\,V_2\,G(z)\,V_3}\ .
\end{equation}
The function $g$ can be expressed in terms of OPE data, which can be seen as follows. Since $g$ has a branch cut from $0$ to $1$, and we can define a function
\begin{equation}
h(z) = \sqrt{z(z-1)}\,g(z)\ .
\end{equation}
Because the square root factor cancels the branch cut, $h$ is a meromorphic function. Furthermore, using the fact that the $V_i$ are super primaries, we see that $h$ has simple poles at $0$ and $1$ and decays as $1/z$ for $z\to \infty$. By Liouville's theorem, it is thus given by the residues at its poles,
\begin{align}
h(z) &= \frac{\text{res}_{u=1}h(u)}{z-1} + \frac{\text{res}_{u=0}h(u)}{z} \nonumber\\
&= \frac{\corr{V_1\,\bbr{G_{0}V_2}\,V_3}}{z-1} + i\,\frac{\corr{V_1\,V_2\,\bbr{G_{0}V_3}}}{z} \ ,
\end{align}
where we have used the OPE
\begin{equation}
G(z)V_j = z^{-\frac{3}{2}}\,G_0\,V_j + z^{-\frac{1}{2}}\,G_{-1}\,V_j + \dots\ ,\quad j=1,2\ ,
\end{equation}
and the factor of $i$ comes from the square root in the definition of $h$.

\begin{verticallines}[black!30!white]
\textbf{Example.}
We now carry out the calculation for the correlator with $w_1=w+1$, $w_2=w$, $w_3=2$. The covering map for this correlator was written down in eq.~(\ref{eq:cov_R_w}) and the moduli $\Theta_i^\mrm{sv}=\Theta_i^\mrm{mv}/\sqrt{z(z-1)}$ are given in eqs.~(\ref{eq:th3_mv}) and (\ref{eq:th2_mv}). The super covering map calculation of the correlator, see eq.~(\ref{eq:3_point_R_susy}), then gives
\begin{equation}
-i\sqrt{\frac{1}{w(w+1)}}\,\theta_3\,\corr{V_1\,\bbr{G_{0}V_2}\,V_3} + (-1)^{\frac{w}{2}}\,\sqrt{\frac{w}{4\,(w+1)}}\,\theta_3\,\corr{V_1\,V_2\,\bbr{G_{0}V_3}}\ .
\end{equation}
Using the reduced covering map in eq.~(\ref{eq:cov_w_R_red}) instead, one obtains
\begin{align}
\theta_3\,\cint{C(0)}\frac{dz}{\sqrt{\partial \Gamma^\text{red}}(z)}\,\frac{h(z)}{\sqrt{z(z-1)}} &= \frac{\theta_3}{\sqrt{-w(w+1)}}\,\cint{C(0)}dz\,\frac{h(z)}{z\bbr{z-1}^{\frac{w}{2}}} \nonumber\\
&= -i\frac{\theta_3}{\sqrt{w(w+1)}}\,\corr{V_1\,\bbr{G_{0}V_2}\,V_3}\nonumber\\
&\quad + (-1)^{\frac{w}{2}}\,\frac{\theta_3}{\sqrt{w(w+1)}}\,\frac{w}{2}\,\corr{V_1\,V_2\,\bbr{G_{0}V_3}}\ .
\end{align}
The two methods of calculation thus give the same result.
\end{verticallines}

\section{The super Liouville action}\label{app:super_liouville}

In this appendix we show how the vacuum correlator on the spherical covering SRS can be evaluated in terms of super covering map data. This leads to the formula eq.~(\ref{eq:super_corr}) for the correlator of super twist fields.

First, we need the supersymmetric analogue of the Weyl anomaly. In the super vielbein formalism \cite{Howe:1978ia}, a super Weyl transformation of some reference vielbein ${E}^A$ is defined as\footnote{Written in the notation of \cite{DHoker:1988pdl}. $M$ denotes all superspacetime indices, $A$ all internal indices, with $a$ the bosonic ones and $\alpha$ the spinor indices.}
\begin{align}
\hat{E}_M^a &= e^{\omega}\,{E}_M ^a\ ,\nonumber\\
\hat{E}_M^\alpha &= e^{\omega/2}\,\left({E}_M^\alpha + {E}_M^a\,(\gamma_a)^{\alpha\beta}{D}_\beta \omega \right)\ .
\end{align}
The super Weyl anomaly can then be written as \cite[eq.~(3.113)]{DHoker:1988pdl}
\begin{equation}\label{eq:superWeyl}
S_{sL}\big(\omega\big) = \frac{c}{96\pi}\,\int d^2\mbf{x}\,\,\mrm{sdet}\big({E}\big)\left({D}\omega\,{\bar{D}}\omega - i\,{\mcl{R}}_{+-}\omega\right)\ ,
\end{equation}
where $d^2\mbf{x}=d^2x\,d\theta\,d\bar{\theta}$ and ${\mcl{R}}$ is the curvature.
Flat superspace is also defined in \cite{Howe:1978ia}. For this vielbein,
\begin{equation}
D = \frac{\partial}{\partial\theta}+\theta\,\frac{\partial}{\partial x}\ ,\quad \bar{D} = \frac{\partial}{\partial\bar{\theta}}+\bar{\theta}\,\frac{\partial}{\partial \bar{x}}\ ,\quad \mcl{R} = 0\ .
\end{equation}

For $\bm{\Gamma}:\,\Sigma_1\to \Sigma_2$ a superconformal map and ${E}^A_1, {E}^A_2$ locally flat vielbeine on $\Sigma_1$ and $\Sigma_2$, respectively, one can calculate the following. The pull-back of ${E}^A_2$ by $\bm{\Gamma}$ to $\Sigma_1$ is related to ${E}^A_1$ by a super Weyl transformation with
\begin{equation}
\omega = \frac{1}{2}\,\log\left(\big({D}\Gamma_2\big)^2\,\big({\bar{D}}\bar{\Gamma}_2\big)^2\right)\ .
\end{equation}
As a sanity check, we can compare this to the bosonic case where $\Gamma_1(\mbf{z}) = f(z)$ with holomorphic $f$. Then $\Gamma_2 = \Theta\,\sqrt{\partial f}$ and $\omega = \frac{1}{2}\,\log\left(\partial f\,\bar{\partial}\bar{f}\right)$ as in \cite{Lunin:2000yv}.

Given these preliminaries, we can now find the super Weyl anomaly for super covering maps, following the derivation of \cite{Lunin:2000yv}. We consider the correlator
\begin{equation}
\corr{\bm{\sigma}_{w_1}\bbr{\mbf{x}_1}\cdots \bm{\sigma}_{w_n}\bbr{\mbf{x}_n}}
\end{equation}
with a sphere covering $\bm{\Gamma}:\mbb{CP}^{1|1}\to \mbb{CP}^{1|1}$ and odd ramifications $w_i$. Denote the ramification points on the covering surface by $\mbf{z}_i$, and the poles by $\bm{\zeta}_j$. By a M\"obius transformation, we can always arrange one of the $\bm{\zeta}_j$ to be $\bbr{\infty,0}$. In the base space, we cut holes around $\mbf{x}_i$ and $\infty$, and similarly around $\mbf{z}_i$ and $\bm{\zeta}_j$ on the covering surface. As the Weyl factor $\omega$ is a sum of a holomorphic and an antiholomorphic function, the super Weyl anomaly eq.~(\ref{eq:superWeyl}) can be partially integrated to give
\begin{equation}
S_{sL}\bbr{\omega} = i\frac{c}{96\pi}\underset{\partial\Sigma}{\int}d\mbf{z}\,\,\omega\, D\omega + \text{c.c.}\ ,
\end{equation}
where $\Sigma$ denotes the covering surface with holes cut out. Near a ramification point $\mbf{z}_i$, the Weyl factor behaves as
\begin{equation}
\omega(\mbf{z}_0) = \frac{1}{2}\log\bbr{w_i a_i\mbf{z}_{0i}^{w_i-1}} + \text{c.c.} + \dots\ ,
\end{equation}
where $a_i$ is the leading coefficient in the Taylor expansion, which can have nilpotent components.
Integrating over the odd variable in the super Liouville action above reduces the contribution from the hole around $\mbf{z}_i$ to
\begin{equation}
i\frac{c}{96\pi}\underset{C(z_i)}{\int}dz\,\,\phi\, \partial \phi + \text{c.c.}\ ,\qquad \phi(z) = \log\bbr{w_i a_i (z-z_i)^{w_i-1}} + \dots\ .
\end{equation}
This is exactly the form in \cite[eq.~(3.8)]{Lunin:2000yv}, and when carrying out the integration one obtains
\begin{equation}
-\frac{c}{24}\frac{w_i-1}{w_i}\log\bbr{a_i} + \text{c.c.} + \text{regulators} \ ,
\end{equation}
where the regulators drop out in the final expressions. The analysis is similar for the pole coefficients: if the behaviour of $\Gamma_1$ near a pole $\bm{\zeta_j}=\bbr{\zeta_j,\varphi_j}$ is
\begin{equation}
\Gamma_1(\mbf{z}) = \frac{C_j}{z-\zeta_j - \Theta\,\varphi_j} + \dots\ ,
\end{equation}
the contribution to the Liouville action is
\begin{equation}
-\frac{c}{12}\log\bbr{C_j} + \text{c.c.} + \text{regulators}\ .
\end{equation}
The pole at $(\infty,0)$ does not contribute to the action.
The total contribution of $\bm{\Gamma}$ to the correlator of twist fields is then given by
\begin{equation}
\corr{\bm{\sigma}_{w_1}\bbr{\mbf{x}_1}\cdots \bm{\sigma}_{w_n}\bbr{\mbf{x}_n}}_{\bm{\Gamma}} = e^{S_{sL}(\omega)}\ ,
\end{equation}
which produces the $a_i,C_j$ dependence shown in eq.~(\ref{eq:super_corr}). The rest of the normalisation can be found by keeping track of the contributions from the holes more carefully, which is completely analogous to \cite{Lunin:2000yv}.

\bibliography{references.bib}

@article{Maldacena:2000hw,
    author = "Maldacena, Juan Martin and Ooguri, Hirosi",
    title = "{Strings in AdS(3) and SL(2,R) WZW model 1.: The Spectrum}",
    eprint = "hep-th/0001053",
    archivePrefix = "arXiv",
    reportNumber = "CALT-68-2245, CITUSC-99-010, HUTP-99-A027, LBNL-44375, UCB-PTH-99-48, LBL-44375",
    doi = "10.1063/1.1377273",
    journal = "J. Math. Phys.",
    volume = "42",
    pages = "2929--2960",
    year = "2001"
}

@article{Eberhardt:2019ywk,
    author = "Eberhardt, Lorenz and Gaberdiel, Matthias R. and Gopakumar, Rajesh",
    title = "{Deriving the AdS$_{3}$/CFT$_{2}$ correspondence}",
    eprint = "1911.00378",
    archivePrefix = "arXiv",
    primaryClass = "hep-th",
    doi = "10.1007/JHEP02(2020)136",
    journal = "JHEP",
    volume = "02",
    pages = "136",
    year = "2020"
}

@article{Eberhardt:2018ouy,
    author = "Eberhardt, Lorenz and Gaberdiel, Matthias R. and Gopakumar, Rajesh",
    title = "{The Worldsheet Dual of the Symmetric Product CFT}",
    eprint = "1812.01007",
    archivePrefix = "arXiv",
    primaryClass = "hep-th",
    doi = "10.1007/JHEP04(2019)103",
    journal = "JHEP",
    volume = "04",
    pages = "103",
    year = "2019"
}

@article{Dei:2020zui,
    author = "Dei, Andrea and Gaberdiel, Matthias R. and Gopakumar, Rajesh and Knighton, Bob",
    title = "{Free field world-sheet correlators for ${\rm AdS}_3$}",
    eprint = "2009.11306",
    archivePrefix = "arXiv",
    primaryClass = "hep-th",
    doi = "10.1007/JHEP02(2021)081",
    journal = "JHEP",
    volume = "02",
    pages = "081",
    year = "2021"
}

@article{Hikida:2020kil,
    author = "Hikida, Yasuaki and Liu, Tianshu",
    title = "{Correlation functions of symmetric orbifold from AdS$_{3}$ string theory}",
    eprint = "2005.12511",
    archivePrefix = "arXiv",
    primaryClass = "hep-th",
    reportNumber = "YITP-20-77",
    doi = "10.1007/JHEP09(2020)157",
    journal = "JHEP",
    volume = "09",
    pages = "157",
    year = "2020"
}

@article{Eberhardt:2021vsx,
    author = "Eberhardt, Lorenz",
    title = "{A perturbative CFT dual for pure NS{\textendash}NS AdS$_{3}$ strings}",
    eprint = "2110.07535",
    archivePrefix = "arXiv",
    primaryClass = "hep-th",
    doi = "10.1088/1751-8121/ac47b2",
    journal = "J. Phys. A",
    volume = "55",
    number = "6",
    pages = "064001",
    year = "2022"
}

@article{Bhat:2021dez,
    author = "Bhat, Faizan and Gopakumar, Rajesh and Maity, Pronobesh and Radhakrishnan, Bharathkumar",
    title = "{Twistor coverings and Feynman diagrams}",
    eprint = "2112.05115",
    archivePrefix = "arXiv",
    primaryClass = "hep-th",
    doi = "10.1007/JHEP05(2022)150",
    journal = "JHEP",
    volume = "05",
    pages = "150",
    year = "2022"
}

@article{Dei:2022pkr,
    author = "Dei, Andrea and Eberhardt, Lorenz",
    title = "{String correlators on $\text{AdS}_3$: Analytic structure and dual CFT}",
    eprint = "2203.13264",
    archivePrefix = "arXiv",
    primaryClass = "hep-th",
    doi = "10.21468/SciPostPhys.13.3.053",
    journal = "SciPost Phys.",
    volume = "13",
    number = "3",
    pages = "053",
    year = "2022"
}

@article{Naderi:2022bus,
    author = "Naderi, Kiarash",
    title = "{DDF operators in the hybrid formalism}",
    eprint = "2208.01617",
    archivePrefix = "arXiv",
    primaryClass = "hep-th",
    doi = "10.1007/JHEP12(2022)043",
    journal = "JHEP",
    volume = "12",
    pages = "043",
    year = "2022"
}

@article{Dei:2023ivl,
    author = "Dei, Andrea and Knighton, Bob and Naderi, Kiarash",
    title = "{Solving AdS$_{3}$ string theory at minimal tension: tree-level correlators}",
    eprint = "2312.04622",
    archivePrefix = "arXiv",
    primaryClass = "hep-th",
    doi = "10.1007/JHEP09(2024)135",
    journal = "JHEP",
    volume = "09",
    pages = "135",
    year = "2024"
}

@article{McStay:2023thk,
    author = "McStay, N. M. and Reid-Edwards, R. A.",
    title = "{Symmetries and covering maps for the minimal tension string on AdS$_{3}${\texttimes} S$^{3}${\texttimes} T$^{4}$}",
    eprint = "2306.16280",
    archivePrefix = "arXiv",
    primaryClass = "hep-th",
    doi = "10.1007/JHEP04(2024)048",
    journal = "JHEP",
    volume = "04",
    pages = "048",
    year = "2024"
}

@article{Knighton:2024qxd,
    author = "Knighton, Bob and Sriprachyakul, Vit",
    title = "{Unravelling AdS$_{3}$/CFT$_{2}$ near the boundary}",
    eprint = "2404.07296",
    archivePrefix = "arXiv",
    primaryClass = "hep-th",
    doi = "10.1007/JHEP01(2025)042",
    journal = "JHEP",
    volume = "01",
    pages = "042",
    year = "2025"
}

@article{Sriprachyakul:2024gyl,
    author = "Sriprachyakul, Vit",
    title = "{Superstrings near the conformal boundary of AdS$_{3}$}",
    eprint = "2405.03678",
    archivePrefix = "arXiv",
    primaryClass = "hep-th",
    doi = "10.1007/JHEP08(2024)203",
    journal = "JHEP",
    volume = "08",
    pages = "203",
    year = "2024"
}

@article{Naderi:2024wqx,
    author = "Naderi, Kiarash",
    title = "{Space-time symmetry from the world-sheet}",
    eprint = "2407.15575",
    archivePrefix = "arXiv",
    primaryClass = "hep-th",
    doi = "10.1007/JHEP03(2025)128",
    journal = "JHEP",
    volume = "03",
    pages = "128",
    year = "2025"
}

@article{Yu:2024kxr,
    author = "Yu, Zhe-fei and Peng, Cheng",
    title = "{Correlators of long strings on AdS$_3\times S^3 \times T^4$}",
    eprint = "2408.16712",
    archivePrefix = "arXiv",
    primaryClass = "hep-th",
    doi = "10.1007/JHEP01(2025)017",
    journal = "JHEP",
    volume = "01",
    pages = "017",
    year = "2025"
}

@article{Knighton:2024pqh,
    author = "Knighton, Bob",
    title = "{Deriving the long-string CFT in AdS$_{3}$}",
    eprint = "2410.16904",
    archivePrefix = "arXiv",
    primaryClass = "hep-th",
    doi = "10.1007/JHEP07(2025)260",
    journal = "JHEP",
    volume = "07",
    pages = "260",
    year = "2025"
}

@article{Yu:2025qnw,
    author = "Yu, Zhe-fei",
    title = "{On the CFT dual of superstring on AdS$_3$}",
    eprint = "2504.20227",
    archivePrefix = "arXiv",
    primaryClass = "hep-th",
    month = "4",
    year = "2025"
}

@article{Eberhardt:2025sbi,
    author = "Eberhardt, Lorenz and Gaberdiel, Matthias R.",
    title = "{A localising AdS$_3$ sigma model}",
    eprint = "2505.09226",
    archivePrefix = "arXiv",
    primaryClass = "hep-th",
    doi = "10.21468/SciPostPhys.19.2.060",
    journal = "SciPost Phys.",
    volume = "19",
    pages = "060",
    year = "2025"
}

@article{Ferreira:2017pgt,
    author = "Ferreira, Kevin and Gaberdiel, Matthias R. and Jottar, Juan I.",
    title = "{Higher spins on AdS$_{3}$ from the worldsheet}",
    eprint = "1704.08667",
    archivePrefix = "arXiv",
    primaryClass = "hep-th",
    doi = "10.1007/JHEP07(2017)131",
    journal = "JHEP",
    volume = "07",
    pages = "131",
    year = "2017"
}

@article{Iguri:2022pbp,
    author = "Iguri, Sergio and Kovensky, Nicolas and Toro, Julian H.",
    title = "{Spectral flow and string correlators in AdS$_3\times S^3 \times T^4$}",
    eprint = "2211.02521",
    archivePrefix = "arXiv",
    primaryClass = "hep-th",
    month = "11",
    year = "2022"
}

@article{Giveon:1998ns,
    author = "Giveon, Amit and Kutasov, D. and Seiberg, Nathan",
    title = "{Comments on string theory on AdS(3)}",
    eprint = "hep-th/9806194",
    archivePrefix = "arXiv",
    reportNumber = "EFI-98-22, RI-4-98, IASSNS-HEP-98-52",
    doi = "10.4310/ATMP.1998.v2.n4.a3",
    journal = "Adv. Theor. Math. Phys.",
    volume = "2",
    pages = "733--782",
    year = "1998"
}

@article{Berkovits:1999im,
    author = "Berkovits, Nathan and Vafa, Cumrun and Witten, Edward",
    title = "{Conformal field theory of AdS background with Ramond-Ramond flux}",
    eprint = "hep-th/9902098",
    archivePrefix = "arXiv",
    reportNumber = "IFT-P-012-99, HUTP-99-A004, IASSNS-HEP-99-5",
    doi = "10.1088/1126-6708/1999/03/018",
    journal = "JHEP",
    volume = "03",
    pages = "018",
    year = "1999"
}

@article{Lunin:2001pw,
    author = "Lunin, Oleg and Mathur, Samir D.",
    title = "{Three point functions for $M^N / S_N$ orbifolds with $\mathcal{N}=4$ supersymmetry}",
    eprint = "hep-th/0103169",
    archivePrefix = "arXiv",
    reportNumber = "OHSTPY-HEP-T-01-005",
    doi = "10.1007/s002200200638",
    journal = "Commun. Math. Phys.",
    volume = "227",
    pages = "385--419",
    year = "2002"
}

@article{Lunin:2000yv,
    author = "Lunin, Oleg and Mathur, Samir D.",
    title = "{Correlation functions for $M^N / S_N$ orbifolds}",
    eprint = "hep-th/0006196",
    archivePrefix = "arXiv",
    reportNumber = "OHSTPY-HEP-T-00-010",
    doi = "10.1007/s002200100431",
    journal = "Commun. Math. Phys.",
    volume = "219",
    pages = "399--442",
    year = "2001"
}

@article{Gaberdiel:2022oeu,
    author = "Gaberdiel, Matthias R. and Nairz, Beat",
    title = "{BPS correlators for AdS$_{3}$/CFT$_{2}$}",
    eprint = "2207.03956",
    archivePrefix = "arXiv",
    primaryClass = "hep-th",
    doi = "10.1007/JHEP09(2022)244",
    journal = "JHEP",
    volume = "09",
    pages = "244",
    year = "2022"
}

@article{Dei:2019iym,
    author = "Dei, Andrea and Eberhardt, Lorenz",
    title = "{Correlators of the symmetric product orbifold}",
    eprint = "1911.08485",
    archivePrefix = "arXiv",
    primaryClass = "hep-th",
    doi = "10.1007/JHEP01(2020)108",
    journal = "JHEP",
    volume = "01",
    pages = "108",
    year = "2020"
}

@article{Pakman:2009ab,
    author = "Pakman, Ari and Rastelli, Leonardo and Razamat, Shlomo S.",
    title = "{Extremal Correlators and Hurwitz Numbers in Symmetric Product Orbifolds}",
    eprint = "0905.3451",
    archivePrefix = "arXiv",
    primaryClass = "hep-th",
    reportNumber = "BROWN-HET-1582, YITP-SB-09-12",
    doi = "10.1103/PhysRevD.80.086009",
    journal = "Phys. Rev. D",
    volume = "80",
    pages = "086009",
    year = "2009"
}

@article{Pakman:2009zz,
    author = "Pakman, Ari and Rastelli, Leonardo and Razamat, Shlomo S.",
    title = "{Diagrams for Symmetric Product Orbifolds}",
    eprint = "0905.3448",
    archivePrefix = "arXiv",
    primaryClass = "hep-th",
    reportNumber = "BROWN-HEP-1573, YITP-SB-09-11",
    doi = "10.1088/1126-6708/2009/10/034",
    journal = "JHEP",
    volume = "10",
    pages = "034",
    year = "2009"
}

@article{Roumpedakis:2018tdb,
    author = "Roumpedakis, Konstantinos",
    title = "{Comments on the S$_{N}$ orbifold CFT in the large $N$-limit}",
    eprint = "1804.03207",
    archivePrefix = "arXiv",
    primaryClass = "hep-th",
    doi = "10.1007/JHEP07(2018)038",
    journal = "JHEP",
    volume = "07",
    pages = "038",
    year = "2018"
}

@inproceedings{Friedan:1986rx,
    author = "Friedan, Daniel",
    title = "{Notes on string theory and two-dimensional conformal field theory}",
    booktitle = "{Workshop on Unified String Theories}",
    reportNumber = "EFI-85-99-CHICAGO",
    editor = "Green, Michael and Gross, David",
    publisher = "World Scientific, Singapore",
    pages = "162--213",
    year = "1986"
}

@article{Witten:2019ylx,
    author = "Witten, Edward",
    title = "{Perturbative superstring theory revisited}",
    doi = "10.4310/PAMQ.2019.v15.n1.a3",
    journal = "Pure Appl. Math. Quart.",
    volume = "15",
    number = "1",
    pages = "213--516",
    year = "2019"
}

@article{Witten:2012ga,
    author = "Witten, Edward",
    title = "{Notes On Super Riemann Surfaces And Their Moduli}",
    eprint = "1209.2459",
    archivePrefix = "arXiv",
    primaryClass = "hep-th",
    doi = "10.4310/PAMQ.2019.v15.n1.a2",
    journal = "Pure Appl. Math. Quart.",
    volume = "15",
    number = "1",
    pages = "57--211",
    year = "2019"
}

@article{Witten:2012bg,
    author = "Witten, Edward",
    title = "{Notes On Supermanifolds and Integration}",
    eprint = "1209.2199",
    archivePrefix = "arXiv",
    primaryClass = "hep-th",
    doi = "10.4310/PAMQ.2019.v15.n1.a1",
    journal = "Pure Appl. Math. Quart.",
    volume = "15",
    number = "1",
    pages = "3--56",
    year = "2019"
}

@article{Rabin:1987rg,
    author = "Rabin, Jeffrey M. and Freund, Peter G. O.",
    title = "{SUPERTORI ARE ALGEBRAIC CURVES}",
    reportNumber = "EFI-87-22-CHICAGO",
    doi = "10.1007/BF01218292",
    journal = "Commun. Math. Phys.",
    volume = "114",
    pages = "131",
    year = "1988"
}

@article{Rabin:1993bw,
    author = "Rabin, Jeffrey M.",
    title = "{Superelliptic curves}",
    eprint = "hep-th/9302105",
    archivePrefix = "arXiv",
    reportNumber = "UCSD-JMR-93-1",
    doi = "10.1016/0393-0440(94)00012-S",
    journal = "J. Geom. Phys.",
    volume = "15",
    pages = "252",
    year = "1995"
}

@article{Distler:1991mf,
    author = "Distler, Jacques and Nelson, Philip C.",
    title = "{Semirigid supergravity}",
    reportNumber = "UPR-0461T, PUPT-1231",
    doi = "10.1103/PhysRevLett.66.1955",
    journal = "Phys. Rev. Lett.",
    volume = "66",
    pages = "1955--1958",
    year = "1991"
}

@article{DHoker:1988pdl,
    author = "D'Hoker, Eric and Phong, D. H.",
    title = "{The Geometry of String Perturbation Theory}",
    reportNumber = "PUPT-1039",
    doi = "10.1103/RevModPhys.60.917",
    journal = "Rev. Mod. Phys.",
    volume = "60",
    pages = "917",
    year = "1988"
}

@article{Howe:1978ia,
    author = "Howe, Paul S.",
    title = "{Super Weyl Transformations in Two-Dimensions}",
    reportNumber = "Print-78-0538 (LANCASTER)",
    doi = "10.1088/0305-4470/12/3/015",
    journal = "J. Phys. A",
    volume = "12",
    pages = "393--402",
    year = "1979"
}

@book{moroianu_2007, place={Cambridge}, series={London Mathematical Society Student Texts}, title={Lectures on Kähler Geometry}, DOI={10.1017/CBO9780511618666}, publisher={Cambridge University Press}, author={Moroianu, Andrei}, year={2007}, collection={London Mathematical Society Student Texts}}

@book{Cavalieri_Miles_2016, place={Cambridge}, series={London Mathematical Society Student Texts}, title={Riemann Surfaces and Algebraic Curves: A First Course in Hurwitz Theory}, publisher={Cambridge University Press}, author={Cavalieri, Renzo and Miles, Eric}, year={2016}, collection={London Mathematical Society Student Texts}}

@book{Varadarajan:2004yz,
author = "Varadarajan, V.~S.",
title = "{Supersymmetry for Mathematicians: An Introduction}",
year = "2004",
publisher = "American Mathematical Society",
collection = "Courant Lecture Notes"
}

@article{Dorrzapf:1997rx,
    author = "Dorrzapf, Matthias",
    title = "{The Definition of Neveu-Schwarz superconformal fields and uncharged superconformal transformations}",
    eprint = "hep-th/9712107",
    archivePrefix = "arXiv",
    reportNumber = "HUTP-97-A054",
    doi = "10.1142/S0129055X99000064",
    journal = "Rev. Math. Phys.",
    volume = "11",
    pages = "137--169",
    year = "1999"
}

@inproceedings{Goddard:1989dp,
    author = "Goddard, Peter",
    title = "{Meromorphic conformal field theory}",
    booktitle = "{Infinite dimensional Lie algebras and Lie groups}",
    reportNumber = "DAMTP-89-01",
    pages = "556--587",
    publisher = "World Scientific, Singapore, New Jersey, Hong Kong",
    editor = "V.~G. Kac",
    year = "1989"
}
\bibliographystyle{JHEP.bst}

\end{document}